\documentclass[twocolumn]{revtex4}
\usepackage[utf8]{inputenc}
\usepackage{graphicx}
\usepackage{color}

\begin{document}

\title{Polarization evolution equation for the exchange-strictionally formed II-type multiferroic materials}

\author{Pavel A. Andreev}
\email{andreevpa@physics.msu.ru}
\affiliation{Department of General Physics, Faculty of physics, Lomonosov Moscow State University, Moscow, Russian Federation, 119991.}

\author{Mariya Iv. Trukhanova}
\email{trukhanova@physics.msu.ru}
\affiliation{Faculty of physics, Lomonosov Moscow State University, Moscow, Russian Federation, 119991.}
\affiliation{Russian Academy of Sciences, Nuclear Safety Institute (IBRAE), B. Tulskaya 52, Moscow, Russian Federation, 115191.}

\date{\today}

\begin{abstract}
The multiferroics are the materials, where single cell of a magnetically order crystal forms an electric dipole moment.
We derive equation for the evolution of the macroscopic density of the electric dipole moment (polarization of the system).
The derivation is made within the "quantum hydrodynamic" method,
which allows to derive equations for the evolution of the macroscopic functions of the quantum systems starting with the microscopic description.
In this work we do not consider the microscopic level of individual electrons and ions,
but we start our analysis from the combination of ions in cells of the crystal.
We present an effective Hamiltonian for the evolution of such intermediate scale objects.
We also apply the equation for the electric dipole moment of the single cell, which is recontracted in the corresponding operator.
Using this operator and the wave function of combination of ions in the cell,
we define the macroscopic density of the electric dipole moment.
Finally, we apply the nonstationary Schrodinger equation with the chosen Hamiltonian in order to derive equation evolution of the polarization.
The derivation is made for the dipole formed by the exchange-striction mechanism
in the II-type multiferroic materials.
The interaction caused term in the polarization evolution equation is found to be proportional to the fourth space derivative of
mixed product (triple scalar product) of three spin density vectors.
Conditions are discussed for the regime, where interaction appears in a smaller order on the space derivatives.
\end{abstract}


\maketitle



\section{Introduction}

Ferromagnetic materials and ferroelectric (seignetoelectric) materials are examples of materials with expressive magnetic or electric properties.
The magneto-electric effect is an example of cooperation of these properties in the single sample.
Redistribution of ions in the crystal cell including ions with the large magnetic moments leads to the formation of the electric dipole moment of the cell.
So, the sample shows the ferroelectric properties.
There are the I-type multiferroic materials,
where the existence of the ferroelectric formation is independent of the ferromagnetic order.
In the II-type multiferroic materials,
the electric dipole moment is formed in relation to the magnetic properties of the material.
There are three mechanisms of the formation of the electric dipole moment in the II-type multiferroics.
The first mechanism is the exchange-striction mechanism arising from the Coulomb exchange interaction modeled
using the Heisenberg Hamiltonian
(which is also called the symmetric spin exchange interaction).
The second mechanism is related to the spin-current model.
It arises from the antisymmetric spin exchange interaction or inverse
Dzyaloshinskii-Moriya model \cite{Khomskii JETP 21}.
Basically, it is related to the exchange part of the spin-orbit interaction.
Let us remind that the spin-orbit interaction is a relativistic effects,
which manifests itself in the weakly-relativistic regime.
From physical point of view, it is the action of the electric field on the moving magnetic moment. 
The third mechanism is the spin dependent p–d hybridization appearing due to the relativistic spin-–orbit coupling.
Each mechanism leads to specific form of the electric dipole moment in relation to the spins/magnetic moments of the ions.
In this work we focus on the first mechanism of dipole moment formation.
We assume that the dipole moment of the cell is given.
So, we do not derive its form,
but we use its form for the definition of the macroscopic polarization.
Our analysis shows that the polarization evolution equation is rather complex.
Therefore, it is essential to consider different interactions separately,
in order to understand their role in the evolution of the multiferroics.
In this paper we focus on the role of the Coulomb exchange interaction
(we use the Heisenberg Hamiltonian).
The Coulomb exchange interaction can contribute in the evolution of polarization of all types of the multiferroics,
but it would required independent consideration of the polarization evolution equation.

There are theoretical works are made in the field of the formation and evolution of polarization in multiferroics.
An effective Hamiltonian model
describing the coupling between the spins and the atomic displacement
(representing the electric polarization) is suggested and applied in Refs.
\cite{Katsura PRL 07}, \cite{Chen EPJB 13} for the spin chirality induced electric polarization.
It appears to be a background for the theory of collective mode dynamics,
which exists in the helical magnets coupled to the electric polarization
via the spin-orbit interaction.
Quasiclassical equations of motion valid for mesoscopic heterostructures with antiferromagnetic order
are obtained in Ref.
\cite{Fyhn PRB 23} based on Keldysh theory for the Green’s functions.
Obtained equations reduce to the
Eilenberger equation and Usadel equation for normal metals
if antiferromagnetism is neglected.
Dynamical multiferroic coupling can be
calculated using density functional theory methods \cite{Juraschek PRM 17},
where multiferroicity is mediated by lattice dynamics, and
the Zeeman splitting of phonon spectra in a magnetic field is analysed.
Magnetic and magnetoelectric excitations in the multiferroic are investigated experimentally in \cite{Pimenov PRL 09}.
Electromagnons and their temperature dependence considered via fitting the excitations by Lorentzian line shapes \cite{Chen PRB 16}.
Equation of evolution of polarization in the system of classical charged particles is derived in terms of classical hydrodynamics
\cite{Drofa TMP 96}.
Similar approach is developed in terms of quantum hydrodynamics,
where the equation of evolution of polarization current is considered as well \cite{Andreev PRB 11}.
Evolution of electrically polarized objects
under simultaneous action of the nonuniform electric and magnetic fields is considered in terms of classical hydrodynamic model
\cite{Andreev EPL 20}, where skyrmions are assumed to posses the electric dipole moment created due to their magnetic structure in chiral magnetic insulators.

In our analysis we use the quantum hydrodynamic method.
Therefore, it is necessary to develop background of the quantum hydrodynamic theory of the ferromagnetic materials,
before some application are presented to the multiferroics.
Required step is made in Ref. \cite{Trukhanova Andreev 2305},
where the contribution of the Heisenberg Hamiltonian in the spin density evolution equations is considered
as the main mechanism of the evolution of the magnetization in ferromagnetic materials on the small time scale intervals,
before dissipation change the character of the evolution.
Let us give some specification of the quantum hydrodynamic method
\cite{Maksimov QHM 99}, \cite{MaksimovTMP 2001}, \cite{Andreev Ch 21},
\cite{Andreev PoF 21}, \cite{Andreev PTEP 19 spin current}, \cite{Shukla RMP 11}.
It is invented as a method of derivation of macroscopic equations for the many-particle quantum systems starting from the microscopic description.
The last one is given by the many-particle Schrodinger equation
(the Pauli equation for the particles with non-zero spin).
The many-particle systems are distributed in the coordinate space.
One of the simplest functions showing this distribution is the particle number density or concentration.
For the magnetized systems it is accompanied with the spin density.
But other functions like the current of particles, the spin current,
and pressure appear as parts of the dynamical equations for the concentration and spin density.
In contrast with the method of the Wigner distribution function \cite{Wigner PR 84}
or the Landau-Silin Fermi-(Bose-)liquid model \cite{Silin ZETF 85}, \cite{Kondratyev LTP 08},
the quantum hydrodynamics deals with the distributions in the coordinate space with no explicit account of the distribution in the momentum space.
Indirectly, the distribution of particles in the momentum space is included
via such functions as the pressure and the thermal part of the many-particle (collective) spin current.

Concerning the physics of the magnetic systems,
the quantum hydrodynamics gives the spin density evolution equation.
It can be considered as a form of the Bloch equation
or the dissipationless form of the
Landau-Lifshitz equation.
The many-particle form of the Bloch equation is considered for the plasmas-like mediums
\cite{MaksimovTMP 2001},
or spinor quantum gases
\cite{Andreev PoF 21}.
In these systems,
in contrast with the crystals with fixed position of atoms and ions,
particles can move,
so the divergence of the collective spin current appears in the hydrodynamic model.
The spin current contains expected convective part $S^{\alpha}v^{\beta}$,
where
$S^{\alpha}$ is the spin density vector,
$v^{\beta}$ is the velocity vector field.
It also includes the thermal part of the spin current,
which is an analog of the thermal pressure for the spin motion \cite{Andreev PTEP 19 spin current}.
If we consider degenerate electron gas we get the Fermi spin current instead of the thermal spin current \cite{Andreev PTEP 19 spin current}.
In this work we consider the exchange interaction as the main mechanism of interaction.
While the exchange interaction is a short-range interaction,
so we follow works
\cite{Andreev LP 19},
\cite{Andreev LP 21},
where method of analysis of the short-range interaction in the quantum hydrodynamics is developed.

This paper is organized as follows.
In Sec. II some features of the physics of the multiferroics required for our analysis is presented.
In Sec. III the background of the quantum hydrodynamic method is presented and the method of its application to the multiferroics is discussed.
In Sec. IV the derivation of the polarization evolution equation is demonstrated.
In Sec. V a brief summary of obtained results is presented.


\section{Modeling of the multiferroics}

We consider the mechanism of the dipole formation in the
II-type multiferroic materials,
which are related to the exchange-striction model.
It can be realized in a system of atoms/ions with parallel spins.
The electric dipole moment of the couple of neighbour atoms appears as a vector proportional to the scalar product of the contributing spins
$\textbf{s}_{i}$ and $\textbf{s}_{j}$ \cite{Tokura RPP 14}
\begin{equation}\label{MFMUEI edm def simm} \textbf{d}_{ij}= \mbox{\boldmath $\Pi$}_{ij} (\textbf{s}_{i}\cdot\textbf{s}_{j}), \end{equation}
where the vector coefficient is introduced to obtain the vector of dipole moment $\textbf{d}_{ij}$.
The dipole moment $\textbf{d}_{ij}$ in the exchange-striction model is formed due to the Coulomb exchange interaction of electrons of neighbouring ions.

Equation (\ref{MFMUEI edm def simm})
contains three constants combined in the single vector constant
$\mbox{\boldmath $\Pi$}_{ij}$.
This constant follow from the Coulomb exchange interaction given by the Heisenberg Hamiltonian.
Being more accurate, we should state
that parameter $\mbox{\boldmath $\Pi$}_{ij}$ is not a constant,
but a function depending on the distance $r_{ij}=\mid \textbf{r}_{i}-\textbf{r}_{j}\mid$
between ions $i$ and $j$.
However, for the chosen phase of the chosen material
we can consider this distance as the fixed value in the further calculations.
Parameter $\mbox{\boldmath $\Pi$}_{ij}=\mbox{\boldmath $\Pi$}_{ij}(r_{ij},\theta_{ij})$ depends on the angle $\theta_{ij}$,
where angle $\theta_{ij}$ is the angle between vector $\textbf{r}_{ij}$ and $z$ axis \cite{Tokura RPP 14}.
We mostly consider radial dependence of function $\mbox{\boldmath $\Pi$}_{ij}(r_{ij},\theta_{ij})$,
but we also consider modification of the model related to the angular dependence.

We can consider evolution of the quasi-classical spins using the equation of classical evolution of magnetic moments.
It can be done using of the methods of classical hydrodynamics.
However, major part of the interactions has the quantum nature,
such as the exchange interaction considered in this paper.
Therefore, we develop the quantum theory of the multiferroics.
Consequently, we need to construct an operator of the electric dipole moment of group of magnetic ions instead of function (\ref{MFMUEI edm def simm}).
Considering spins $\textbf{s}_{i}$ as the corresponding operator.
Moreover, we consider a model example of spin-1/2 ions.
This is an unnecessary assumption,
so the model can be rederived for other spin values.
Hence, we introduce the operator of the electric dipole moment
\begin{equation}\label{MFMUEI edm operator}
\hat{\textbf{d}}_{ij}= \mbox{\boldmath $\Pi$}_{ij} (\hat{\textbf{s}}_{i}\cdot\hat{\textbf{s}}_{j}), \end{equation}
with commutator
\begin{equation}\label{MFMUEI commutator of spins}
[\hat{s}_{i}^{\alpha},\hat{s}_{j}^{\beta}]=\imath\hbar\delta_{ij}\varepsilon^{\alpha\beta\gamma} \hat{s}_{i}^{\gamma}, \end{equation}
where
$\alpha$, $\beta$, $\gamma$ are the tensor indexes,
so each of them is equal to $x$, $y$, $z$,
summation on the repeating Greek symbol is assumed,
$\imath$ is the imaginary unit $\imath^{2}=-1$,
$\delta_{ij}$ is the three-dimensional Kronecker symbol,
$\varepsilon^{\alpha\beta\gamma}$ is the three-dimensional Levi-Civita symbol.
Since we consider spin-1/2 ions
we can represent the spin operator $\hat{\textbf{s}}_{i}$ via the Pauli matrixes.
Below, we consider the quantum average of the presented operator (\ref{MFMUEI edm operator}) on the many-particle wave function,
in accordance with the quantum hydrodynamic method.


Operator (\ref{MFMUEI edm operator}) refers to the couple of particles,
but they are neighbour particles.
We need to include this properties before
we start construction of the hydrodynamic equations.
If particles are placed in line with fixed positions,
we need to consider particles with numbers $i$ and $j=i+1$.
Instead of numeration of particles in sequence
(or three sequences for the three dimensional samples)
we want to focus on the relative space position of the interacting particles.
To this goal,
we represent definition (\ref{MFMUEI edm operator}) in the following form
\begin{equation}\label{MFMUEI edm operator Mod}
\hat{\textbf{d}}_{i}=\sum_{j\neq i}
\mbox{\boldmath $\Pi$}_{ij}(r_{ij}) (\hat{\textbf{s}}_{i}\cdot\hat{\textbf{s}}_{j}), \end{equation}
where $\mbox{\boldmath $\Pi$}_{ij}(r_{ij})$ is a function,
which has fast decrease to the zero value at the increase of the distance between particles.
We can define it explicitly via its asymptotics
$\mbox{\boldmath $\Pi$}_{ij}(r_{ij})=\mbox{\boldmath $\Pi$}_{ij}$ at $r< a_{s}$ 
$\mbox{\boldmath $\Pi$}_{ij}(r_{ij})=0$ at $r> a_{l}$, 
where on the right-hand side of one of equations we use constant vector
$\mbox{\boldmath $\Pi$}_{ij}$ introduced in equations
(\ref{MFMUEI edm def simm}) and (\ref{MFMUEI edm operator}).
Parameters $a_{s}$ and $a_{l}$ are the characteristic small $a_{s}$ and large $a_{l}$ distances.
Spherical layer between $a_{s}$ and $a_{l}$ is an intermediate interval of physically smooth transition between two regimes.
We mentioned above that constant vector $\mbox{\boldmath $\Pi$}_{ij}$
depends on the angle,
but we neglect it during major calculations.
However, we estimate the contribution of the angle dependence
to understand possibility of generalizations.
Therefore, parameters $a_{s}$ and $a_{l}$ can be also considered as the functions of direction,
but at this step of the model development it is an unnecessary complication.
Let us repeat that parameter $\mbox{\boldmath $\Pi$}_{ij}$ introduced in equation (\ref{MFMUEI edm def simm})
is a function of the distance of two neighboring ions forming the dipole moment.
But this distance is assumed to be fixed.
Here, we try to distinguish neighboring ions and ions located further.
So, distant ions do not give contribution in the dipole moment formation.
As we can see,
the space interval between asymptotics
is a small intermediate area,
which can be modeled in a simple way in order to estimate its contribution.
We can assume $a_{l}=a_{s}$,
so function $\mbox{\boldmath $\Pi$}_{ij}(r_{ij})$ is modeled as the Heaviside step function.
Below, we refer to it as an effective function with a short radius of its action.

\section{Quantum hydrodynamics}

The derivation of hydrodynamic model is usually starts with the definition of the concentration/number density of particles
\cite{MaksimovTMP 2001}, \cite{Andreev PoF 21}.
It shows us the distribution of particles in the coordinate space.
If we can assume
that the distribution of particles has negligible change,
we need to choose another function to start the derivation.
We consider the spin density $\textbf{S}(\textbf{r},t)$,
which appears at the hydrodynamic description of particles with spin anyway \cite{MaksimovTMP 2001}, \cite{Andreev PoF 21}
\begin{equation}\label{MFMUEI S def} \textbf{S}(\textbf{r},t)=
\int \Psi_{S}^{\dagger}(R,t)\sum_{i}\delta(\textbf{r}-\textbf{r}_{i})
(\hat{\textbf{s}}_{i}\Psi(R,t))_{S}dR. \end{equation}
This definition is quite straightforward.
Here, we have operators of spin of elements of our system $\hat{\textbf{s}}_{i}$.
Therefore, structure
$\hat{\textbf{S}}=\sum_{i}\delta(\textbf{r}-\textbf{r}_{i}) \hat{\textbf{s}}_{i}$
represents the operator of the spin density of whole system.
Here, coordinate $\textbf{r}_{i}$ is the effective coordinate of our quasi-point-like objects.
Hence, the spin density definition has the following structure
$\textbf{S}=\int\Psi^{\dagger}\hat{\textbf{S}}\Psi dV$,
which resemblance the quantum average of the chosen operator, giving the observable value of this function,
where $dV\equiv dR$ is the element of the configurational space.
The multi-spinor wave function $\Psi(R,t))_{S}$ is the wave function of whole system.
Here we consider composed objects,
which are combinations of ions in the cell of crystal.
We do not consider the wave function for the electrons and ions,
which would give a microscopic level of description.
We work on the large space scale of composed objects,
which are modeled as the point-like quasi-particles.
Argument of the wave function $\Psi(R,t))_{S}$ is the
assemblage of coordinates of all quasi-particles
$R=\{\textbf{r}_{1},...,\textbf{r}_{N}\}$,
where $N$ is the number of quasi-particle elements of our system
and $\textbf{r}_{i}$ is the coordinate of $i$-th element.
Index of the wave function $\Psi(R,t))_{S}$ is the
assemblage of spin values of all elements $S=\{s_{1}, ..., s_{N}\}$.

Definition (\ref{MFMUEI S def}) shows
that the evolution of the spin density of the system depends on the evolution of the wave function.
Hence, we need to use the many-particle Schrodinger/Pauli equation
$\imath\hbar\partial_{t}\Psi_{S}=\hat{H}_{SS'}\Psi_{S'}$
with a Hamiltonian containing main interactions of elements of our system.

In this work we focused on the Coulomb exchange interaction in terms of the
Heisenberg Hamiltonian:
$\hat{H}_H=-J \hat{\textbf{s}}_{1}\cdot\hat{\textbf{s}}_{2}$,
where $\hat{\textbf{s}}_{i}$ are the spins of two interacting electrons.

First, we need to point out the area of applicability of this Hamiltonian.
If we calculate average energy of the interaction of two electrons
(or other spin-1/2 particles)
we find that this interaction is composed of two terms.
This picture appears
in the weak interaction limit,
so the wave function can be presented as combination of the single particle wave functions.
We interested in the interaction of valence electrons of atoms or ions.
One part of average interaction can be associated with the classical structure
which is expressed via the square module of the wave function of each electron.
Second part of the interaction term called the exchange interaction,
which is constructed of the single-particle wave functions
while these wave functions do not combine in any observable.
Therefore, it do not have any quasi-classical interpretation.
Depending on the full spin of two interacting electrons
we have different symmetry for the coordinate space part of the two-particle wave function defining obtained integrals for the average energy.
Full spin can have two values $0$ or $1$.
Hence, the spin-dependent contribution of the exchange interaction can be presented via operators of spin of each electron
$\hat{\textbf{s}}_{1}$, and $\hat{\textbf{s}}_{2}$
in the well-known form of
Heisenberg Hamiltonian:
$\hat{H}_H=-J \hat{\textbf{s}}_{1}\cdot\hat{\textbf{s}}_{2}$.

Described analysis is correct for the couple of electrons or couple of atoms with one valence electron,
where the exchange interaction mainly corresponds to the interaction of the valence electrons.
If we consider the interaction of atoms or ions with more complex structure of the valence electrons
we get more complex result for the exchange interaction.
For instance, we have two valence electrons in each atom,
so we have interaction of four electrons.
For such system, in general case,
there is no separation of the wave function on the coordinate and spin parts.
We can try to introduce a rough approximation,
where we assume
that we have interaction of two "black boxes".
So, we assume that each "box" has spin equal to $1$.
Moreover, it allows us to exclude the exchange interaction of pairs of electrons inside each atom.
Since, we need to obtain an effective interaction between atoms.

Consequently, we consider a pair of spin-1 objects.
We can introduce two-particle wave function presented as the product of the coordinate and spin wave functions.
Moreover, full spin of this system can have three values $0$, $1$, or $2$.
In order to obtain the spin depending operator to describe different values of the exchange interaction for the different full spins of the system
we need to use the superposition of two operators
$\hat{\textbf{s}}_{1}\cdot\hat{\textbf{s}}_{2}$ and $(\hat{\textbf{s}}_{1}\cdot\hat{\textbf{s}}_{2})^{2}$.
Particularly, extended Heisenberg Hamiltonian is used in the spinor quantum atomic gases \cite{Kawaguchi Ph Rep 12}, \cite{Stamper-Kurn RMP 13}.

Term $\hat{\textbf{s}}_{1}\cdot\hat{\textbf{s}}_{2}$ exists for any value of spin of interacting objects.
Hence, we consider its contribution for different systems with no specification of the spin.

In more general case we can consider Hamiltonian in the following form
$\hat{H}=\sum_{i=1}^{N}[\hat{\textbf{p}}^{2}_{i}/2m$
$+V_{ext}(\textbf{r}_{i},t)$
$-(1/2)\sum_{j=1, j\neq i}^{N} U(\textbf{r}_{i}-\textbf{r}_{j})
(\hat{\textbf{s}}_{i}\cdot\hat{\textbf{s}}_{j}) ]$.
Here we have the kinetic energy $\sum_{i=1}^{N}(\hat{\textbf{p}}^{2}_{i}/2m$,
which can be dropped since considering elements do not involve in the translational motion,
and the thermal oscillations of ions combining these elements can be dropped till it become necessary to generalize suggested model.
Hamiltonian $\hat{H}$ also includes the potential of the external fields $V_{ext}(\textbf{r}_{i},t)$.
The last term corresponds to the interparticle interaction or in our case it is interaction of combined elements described above.
This interaction is written as an effective spin-spin interaction.
Depending on the form and properties of function $U$ we can consider various interactions.
As main examples we can point out the dipole-dipole interaction
and the
Heisenberg Coulomb exchange interaction.
Here, we can include the spin-orbit interaction or its effective exchange part.
This structure can include the Zeeman energy $-\mu_{i}(\hat{\textbf{s}}_{i}\cdot \textbf{B}_{i})$
as a part of the external field potential $V_{ext}(\textbf{r}_{i},t)$.

Finally, we restrict ourselfs with the account of the single term in Hamiltonian corresponding
to the Heisenberg Coulomb exchange interaction
\begin{equation}\label{MFMUEI H Ham short} \hat{H}= -\frac{1}{2}\sum_{i=1}^{N}\sum_{j=1, j\neq i}^{N} U(\textbf{r}_{i}-\textbf{r}_{j})
(\hat{\textbf{s}}_{i}\cdot\hat{\textbf{s}}_{j}), \end{equation}
where
$U(\textbf{r}_{i}-\textbf{r}_{j})$ corresponds to the space dependence of the exchange integral $J$.

So, if we want to consider the spin density evolution (\ref{MFMUEI S def})
we calculate the time derivative of the definition (\ref{MFMUEI S def}).
The time derivative acts on the wave functions under the integral.
We replace these derivatives of the wave functions with the Hamiltonian in accordance with the many-particle Schrodinger/Pauli equation
with Hamiltonian (\ref{MFMUEI H Ham short}).
After some technical calculations discussed in Ref. \cite{Trukhanova Andreev 2305}
we obtain the dissipationless form of the Landau-Lifshitz equation
\begin{equation}\label{MFMUEI LL eq dissipless}
\partial_{t}\textbf{S}=
\frac{1}{6}g_{u}
[\textbf{S}, \triangle\textbf{S}], \end{equation}
where coefficient
$g_{u}=\int \xi^{2}U(\xi) d\mbox{\boldmath $\xi$}$
appears as an integral of the  exchange integral over a small area of space,
where this integral has nonzero value.
For the model examples of $U(r)$ this integral can be calculated to extract nonintegral relation between coefficient in equation
(\ref{MFMUEI LL eq dissipless}) and an effective value of the exchange integral $J$.
Moreover, the Zeeman energy can provide additional well-known mechanism of the evolution of the spin density \cite{MaksimovTMP 2001}
$\partial_{t}\textbf{S}=
\frac{2\mu_{a}}{\hbar}
[\textbf{S}, \textbf{B}],$
where
$\mu_{a}$ is the magnetic moment of particle of chosen material,
if we consider the electron gas it would be equal to $\mu_{a}=-\mid\mu_{a}\mid$,
with $\mu_{a}$ is given by Bohr magneton.
We stress our attention on the contribution of the exchange interaction in the spin density evolution equation
since we consider the Heisenberg Coulomb exchange interaction in the polarization evolution equation in other sections below.
It is a part of the Landau-–Lifshitz–-Gilbert
discussed in Ref. \cite{Barman JAP 20} in context of the magnetization dynamics of nanoscale magnetic materials.

%

\section{Electric polarization evolution}

In this section we present the derivation of the polarization evolution equation
using the quantum hydrodynamic method.
To this goal, we present the macroscopic definition of the polarization/electric dipole moment density
as the average of the corresponding operator (\ref{MFMUEI edm operator Mod}) on the many-particle wave function $\Psi_{S}(R,t)$:
\begin{equation}\label{MFMUEI P def} \textbf{P}(\textbf{r},t)=
\int \Psi_{S}^{\dagger}(R,t)\sum_{i}\delta(\textbf{r}-\textbf{r}_{i})
(\hat{\textbf{d}}_{i}\Psi(R,t))_{S}dR, \end{equation}
where
operator of the electric dipole moment $\hat{\textbf{d}}_{i}$ is given by equation (\ref{MFMUEI edm operator Mod}).
We also use the notation $R\equiv R_{N}$ for the coordinates of all particles in the $3N$-dimensional configurational space.
In equation (\ref{MFMUEI P def}) we drop subindex $R$
since here and all other major equations we have same number of particles.
However, if we consider subsystems of smaller numbers of particles we specify the number of particles like $R_{N-1}$ or $R_{N-2}$.
To be more precise the spin density (\ref{MFMUEI S def})
or polarization (\ref{MFMUEI P def}) should be presented in the symmetric form
$\textbf{P}(\textbf{r},t)=(1/2)
\int \sum_{i}\delta(\textbf{r}-\textbf{r}_{i})\{\Psi_{S}^{\dagger}(R,t)
(\hat{\textbf{d}}_{i}\Psi(R,t))_{S}+c.c.\}dR$,
where $h.c.$ stands for the Hermitian conjugation.

\subsection{General structure of equations}

Start of the derivation of the polarization evolution equation is relatively simple.
We consider the time derivative of the polarization vector function (\ref{MFMUEI P def}).
This derivative acts on the wave function being under the integral.
Time derivative of each wave function can be replaced by the action of the Hamiltonian operator (\ref{MFMUEI H Ham short}).
It gives expression for the time derivative of the polarization
via the commutator of the Hamiltonian operator and the electric dipole moment operator
\begin{equation}\label{MFMUEI P time der via H} \partial_{t}\textbf{P}(\textbf{r},t)=
\frac{\imath}{\hbar}\int \Psi^{\dagger}(R,t)\sum_{i}\delta(\textbf{r}-\textbf{r}_{i})
[\hat{H},\hat{\textbf{d}}_{i}]\Psi(R,t)dR, \end{equation}
where we dropped the spin subindex of the wave function to simplify presenting equations.
We use explicit result of calculation of the commutator
$[\hat{H},\hat{\textbf{d}}_{i}]$
using operators (\ref{MFMUEI edm operator Mod}) and (\ref{MFMUEI H Ham short}).
It leads to the following expression
$[\hat{H},\hat{d}_{i}^{\alpha}]=$
$-\imath\hbar\varepsilon^{\beta\gamma\delta}
\sum_{j\neq i}\sum_{n\neq i,j}\Pi_{ij}^{\alpha}(U_{ni}-U_{nj})
\hat{s}_{n}^{\beta}\hat{s}_{j}^{\gamma}\hat{s}_{i}^{\delta}$.
It leads to the following expression for the polarization evolution
$$\partial_{t}\textbf{P}(\textbf{r},t)=
\varepsilon^{\beta\gamma\delta}\int \sum_{i,j,i\neq j}\sum_{n\neq i,j}\delta(\textbf{r}-\textbf{r}_{i})\times$$
\begin{equation}\label{MFMUEI P time der via expl form of comm}
\times\Pi_{ij}^{\alpha}(U_{ni}-U_{nj})
\Psi^{\dagger}(R,t)\hat{s}_{n}^{\beta}\hat{s}_{j}^{\gamma}\hat{s}_{i}^{\delta}
\Psi(R,t)dR. \end{equation}
This equation is the general expression,
where we need to include information on properties of functions $\Pi_{ij}^{\alpha}$ and $U_{ij}$.
These functions decreases to zero at the relatively small distances between objects $i$ and $j$.

Roughly estimating the found structure
we see
that it is composed of three spin operators related to three different particles.
Hence, the final expression is composed of the mixed product (triple scalar product) of the spin-density vectors
with possible space derivatives of the spin density
(since the spins itself gives the zero value product).
We can conclude
that most simple nonzero combination of the spin vectors and the space derivatives
$\partial_{t}P^{\alpha}\sim (\textbf{S}\cdot [\nabla^{\beta}\textbf{S}\times \triangle\textbf{S}])$.
However, below we show that this combination give no contribution in the polarization evolution
since coefficient in front of it is equal to zero.
Hence, the nonzero contribution appears from the combination with one additional space derivative:
\begin{equation}\label{MFMUEI P time der estimation}
\partial_{t}P^{\alpha}\sim
\nabla^{\beta}
(\textbf{S}\cdot [\nabla^{\beta}\textbf{S}\times \triangle\textbf{S}]), \end{equation}
where
$\nabla^{\beta}$ is the
gradient vector operator composed of the space coordinate derivatives,
$\triangle=\nabla^{2}=\nabla^{\beta}\nabla^{\beta}$ is the laplace operator.
Here, we state that direction of the polarization $P^{\alpha}$ differs from the direction gradient operator $\nabla^{\beta}$.
This relation constructs via a tensor coefficient.
This coefficient would depend on the microscopic characteristics of the systems existing in equation (\ref{MFMUEI P time der via expl form of comm}).
These are $\Pi_{ij}^{\alpha}$ and $U_{ni}$.
Below, we consider three items:
\newline
1) Microscopic justification of predicted expression
(including the proof of account of additional space derivative in the mean-field approximation).
\newline
2) Derivation of the explicit form of coefficient.
\newline
3) Finding conditions for the applicability of predicted expression,
since it is based on the simple assumption that operators of spin would give most obvious representation via the spin density.

We have three ions $i$, $j$ and $n$.
Two of them $i$, $j$ are related to the formation of the electric dipole moment of the cell containing these ions.
We can see it due to the presence of function $\Pi_{ij}^{\alpha}$ with corresponding indexes.
Each of these two ions interact with surrounding ions via the pair-potential $U_{ni}$.
Therefore, it is unnecessary to consider all surrounding ions,
but it is possible to focus on the single ion.
Equation (\ref{MFMUEI P time der via H}) is composed of two terms,
since we have a difference of two potentials $(U_{ni}-U_{nj})$.
We can consider these terms separately or simultaneously.
It has no specific difference from the methodological point of view.
In both cases we need to consider three particles $i$, $j$ and $n$.
We also need to consider their relative positions,
which should be small since two neigboring ions form electric dipole moment
and they interact with the closest neighbours.
To this goal we consider the coordinate of center of mass of three particle
$\textbf{R}_{ijn}=(\textbf{r}_{i}+\textbf{r}_{j}+\textbf{r}_{n})/3$
and two relative coordinates
$\textbf{r}_{in}\equiv\textbf{r}_{1}=\textbf{r}_{i}-\textbf{r}_{n}$
and
$\textbf{r}_{jn}\equiv\textbf{r}_{2}=\textbf{r}_{j}-\textbf{r}_{n}$.
The relative coordinate of the third pair is easily represented via introduced relative distances
$\textbf{r}_{ij}\equiv\textbf{r}_{3}=\textbf{r}_{1}-\textbf{r}_{2}$.
Let us also show reverse expressions for the coordinates of each particle via relative coordinates and the coordinate of the center of mass:
$\textbf{r}_{i}=\textbf{R}_{ijn}+(2/3)\textbf{r}_{in}-(1/3)\textbf{r}_{jn}$,
$\textbf{r}_{j}=\textbf{R}_{ijn}-(1/3)\textbf{r}_{in}+(2/3)\textbf{r}_{jn}$,
and
$\textbf{r}_{n}=\textbf{R}_{ijn}-(1/3)\textbf{r}_{in}-(1/3)\textbf{r}_{jn}$.

In equation (\ref{MFMUEI P time der via expl form of comm}),
we have contribution of coordinates $\textbf{r}_{i}$, $\textbf{r}_{j}$, $\textbf{r}_{n}$
in the delta-function $\delta(\textbf{r}-\textbf{r}_{i})$
and in the wave function $\Psi(R,t)=\Psi(...,\textbf{r}_{i}, ..., \textbf{r}_{j}, ..., \textbf{r}_{n}, ...,t)$.
We need to make expansion of these functions on the small relative distances $\textbf{r}_{in}$ and $\textbf{r}_{jn}$.

\subsection{Zeroth order expansion}

After expansion under integral in equation (\ref{MFMUEI P time der via expl form of comm})
on the small value of the relative distances and consideration of the zeroth order on the small parameter
we find the following consequence of equation (\ref{MFMUEI P time der via expl form of comm})
$$\partial_{t}P^{\alpha}(\textbf{r},t)=
\varepsilon^{\beta\gamma\delta}\int \sum_{i,j,i\neq j}\sum_{n\neq i,j}\delta(\textbf{r}-\textbf{R}_{ijk})
\Pi_{ij}^{\alpha}\times$$
$$\times(U_{ni}-U_{nj})\Psi^{\dagger}(...,\textbf{R}_{ijk}, ..., \textbf{R}_{ijk}, ..., \textbf{R}_{ijk}, ...)\times$$
\begin{equation}\label{MFMUEI P time der 0 order}
\times
\hat{s}_{n}^{\beta}\hat{s}_{j}^{\gamma}\hat{s}_{i}^{\delta}
\Psi(...,\textbf{R}_{ijk}, ..., \textbf{R}_{ijk}, ..., \textbf{R}_{ijk}, ...)dR. \end{equation}

In order to simplify the notation during more complex equations demonstrated below
we make the following representation
$\hat{s}_{n}^{\beta}\hat{s}_{j}^{\gamma}\hat{s}_{i}^{\delta}
\Psi(...,\textbf{R}_{ijk}, ..., \textbf{R}_{ijk}, ..., \textbf{R}_{ijk}, ...)$
$\equiv\hat{s}_{3}^{\beta}\hat{s}_{2}^{\gamma}\hat{s}_{1}^{\delta}
\Psi(\textbf{R}_{1}, ..., \textbf{R}_{2}, ..., \textbf{R}_{3})$
$\equiv\Psi(\textbf{R}_{1}, \textbf{R}_{2}, \textbf{R}_{3}, R_{N-3})$,
where $\textbf{R}_{1}=\textbf{R}_{2}=\textbf{R}_{3}=\textbf{R}_{ijk}$,
but we keep subindexes "1", "2" and "3" to point out argument of the many-particle wave function.
Notation $R_{N-3}$ is used for the coordinates of remaining particles of the system.
Replacement of the coordinates of $i$, $j$ and $n$ particles on the first, the second and the third place is not assumed here.

Let us repeat equation (\ref{MFMUEI P time der 0 order}) including simplified notations
$$\partial_{t}P^{\alpha}(\textbf{r},t)=
\varepsilon^{\beta\gamma\delta}\int \sum_{i,j,i\neq j}\sum_{n\neq i,j}\delta(\textbf{r}-\textbf{R}_{ijk})
\Pi_{ij}^{\alpha}(U_{ni}-U_{nj})\times$$
\begin{equation}\label{MFMUEI P time der 0 order simple notation}
\times\Psi^{\dagger}(\textbf{R}_{1}, \textbf{R}_{2}, \textbf{R}_{3}, R_{N-3})
\hat{s}_{n}^{\beta}\hat{s}_{j}^{\gamma}\hat{s}_{i}^{\delta}
\Psi(\textbf{R}_{1}, \textbf{R}_{2}, \textbf{R}_{3}, R_{N-3})dR. \end{equation}

An important property of equation (\ref{MFMUEI P time der 0 order}) is the separation of variables of relative motion from the coordinate of the center of mass and coordinates of other $N-3$ particles.
So, we can introduce an interaction constant:
\begin{equation}\label{MFMUEI int const 0 order}
g_{0}^{\alpha}=\int d\textbf{r}_{1}d\textbf{r}_{1} \Pi^{\alpha}(\mid \textbf{r}_{1}-\textbf{r}_{2}\mid)(U(r_{1})-U(r_{2})),
\end{equation}
where $r_{1}=\mid\textbf{r}_{1}\mid$,
$d\textbf{r}_{1}\equiv d^{3}r_{1}$, etc.
Separation of interaction constants happens in each order of the expansion.
To make this separation we need to notice
$dR=d\textbf{r}_{i}d\textbf{r}_{j}d\textbf{r}_{n}dR_{N-3}=d\textbf{r}_{ni}d\textbf{r}_{nj}d\textbf{R}_{ijn}dR_{N-3}$.

After extraction of the interaction constant
we need to deal with the integral over the center of mass and integrals over the coordinates of other particles.
They are related via the correlations in the many-particle wave function.
Considering coordinates are related to the "core" of atoms/ions composed of the nucleus and inner electrons
(all electrons except the valence electrons).
So, these coordinates can be roughly associated with the nucleuses.
The derivation of the Landau- Lifshitz equation
(\ref{MFMUEI LL eq dissipless}) shows
that correlation between nucleuses can be dropped in order to derive well-known macroscopic result.
Consequently, we can apply same approximation for the polarization evolution in the multiferroic material.
Hence, we use the following expansion of the many-particle wave function
$$\Psi(\textbf{R}_{1}, \textbf{R}_{2}, \textbf{R}_{3}, R_{N-3})\equiv
\langle \textbf{R}_{1}, \textbf{R}_{2}, \textbf{R}_{3}, R_{N-3} | n_{1}, n_{2}, ..., n_{f}, ... \rangle$$
$$=\sum_{f}\sqrt{\frac{n_{f}}{N}}\langle \textbf{R}_{1}| f\rangle\langle \textbf{R}_{2}, \textbf{R}_{3}, R_{N-3} | n_{1}, n_{2}, ..., n_{f}-1, ... \rangle$$
$$=\sum_{f}\sum_{f'\neq f}\sqrt{\frac{n_{f}n_{f'}}{N(N-1)}}
\langle \textbf{R}_{1}| f\rangle \langle \textbf{R}_{2}| f'\rangle\times$$
$$\times
\langle \textbf{R}_{3}, R_{N-3} | n_{1}, n_{2}, ..., n_{f}-1, ..., n_{f'}-1, ... \rangle
=\sum_{f}$$
$$\sum_{f'\neq f}\sum_{f''\neq f,f'}\sqrt{\frac{n_{f}n_{f'}n_{f''}}{N(N-1)(N-2)}}
\langle \textbf{R}_{1}| f\rangle \langle \textbf{R}_{2}| f'\rangle \langle \textbf{R}_{3}| f''\rangle\times$$
\begin{equation}\label{MFMUEI WF expnsion 3}
\times \langle R_{N-3} | n_{1}, n_{2}, ..., n_{f}-1, ..., n_{f'}-1, ..., n_{f''}-1, ... \rangle.
\end{equation}
We also need the orthogonalization condition for function
$\langle R_{N-3} | n_{1}, n_{2}, ..., n_{f}-1, ..., n_{f'}-1, ..., n_{f''}-1, ... \rangle$:
$$\langle R_{N-3} | n_{1}, n_{2}, ..., n_{g}-1, ..., n_{g'}-1, ..., n_{g''}-1, ... \rangle^{\dag}\times$$
$$\times\langle R_{N-3} | n_{1}, n_{2}, ..., n_{f}-1, ..., n_{f'}-1, ..., n_{f''}-1, ... \rangle$$
\begin{equation}\label{MFMUEI WF orthogonalization}
=\delta_{fg}\delta_{f'g'}\delta_{f''g''}.
\end{equation}
If we want to include the exchange correlations we can find more accurate expansion in Ref. \cite{Maksimov QHM 99}.
It is also discussed in Ref. \cite{Andreev LP 21}.
Here we use single particle wave functions
$\phi_{f}(\textbf{R}_{1})\equiv\langle \textbf{R}_{1}| f\rangle$.

Final interpretation of equation (\ref{MFMUEI P time der 0 order simple notation})
requires corresponding representation of the spin density (\ref{MFMUEI S def}).
The spin density (\ref{MFMUEI S def}) contains one operator corresponding to the $i$-th particle.
Hence we can use the following expansion of the wave function
$$\Psi(\textbf{r}_{i}, R_{N-1})\equiv
\langle \textbf{r}_{i}, R_{N-1} | n_{1}, n_{2}, ..., n_{f}, ... \rangle$$
\begin{equation}\label{MFMUEI WF expnsion 1 i}
=\sum_{f}\sqrt{\frac{n_{f}}{N}} \langle \textbf{r}_{i}| f\rangle
\langle R_{N-1} | n_{1}, n_{2}, ..., n_{f}-1, ... \rangle .
\end{equation}
Using the approximate single particle form we obtain the following expression for the spin density (\ref{MFMUEI S def}):
$\textbf{S}=\sum_{f}n_{f}\phi_{f}(\textbf{r})^{\dag}\hat{\textbf{s}}\phi_{f}(\textbf{r})$.
We use this notations to complete the calculation of expression (\ref{MFMUEI P time der 0 order simple notation})
presenting it in terms of the spin densities:
\begin{equation}\label{MFMUEI P time der 0 order fin}
\partial_{t}\textbf{P}(\textbf{r},t)=g_{0}^{\alpha}
\varepsilon^{\beta\gamma\delta} S^{\beta}(\textbf{r},t)S^{\gamma}(\textbf{r},t)S^{\delta}(\textbf{r},t) =0,
\end{equation}
where we use $\textbf{R}_{1}=\textbf{R}_{2}=\textbf{R}_{3}=\textbf{R}_{ijk}$
and projection of $\textbf{R}_{ijk}$ on the three-dimensional space $\textbf{r}$ via the delta-function.
Presented expression (\ref{MFMUEI P time der 0 order fin}) is equal to zero due to the vector product of identical vectors
written via the Levi-Civita symbol.

\subsection{First order expansion}

In this subsection we use a part of equation (\ref{MFMUEI P time der via expl form of comm})
appearing in the first order of the expansion on the small interparticle distances.

It can be written as two group of terms
$\textbf{P}_{(1)}(\textbf{r},t)=\textbf{P}_{(1a)}(\textbf{r},t)+\textbf{P}_{(1b)}(\textbf{r},t)$,
where
$$\partial_{t}P^{\alpha}_{(1a)}(\textbf{r},t)=\frac{1}{3}
\varepsilon^{\beta\gamma\delta}\nabla\int \sum_{i,j,i\neq j}\sum_{n\neq i,j}
\delta(\textbf{r}-\textbf{R}_{ijk})\times$$
\begin{equation}\label{MFMUEI P time der I order a}
\times(\textbf{r}_{jn}-2\textbf{r}_{in})\Pi_{ij}^{\alpha}(U_{ni}-U_{nj})\Psi^{\dagger}(R',t)\hat{s}_{n}^{\beta}\hat{s}_{j}^{\gamma}\hat{s}_{i}^{\delta}
\Psi(R',t)dR, \end{equation}
and
$$\partial_{t}P^{\alpha}_{(1b)}(\textbf{r},t)=\frac{1}{3}
\varepsilon^{\beta\gamma\delta}\int \sum_{i,j,i\neq j}\sum_{n\neq i,j}\delta(\textbf{r}-\textbf{R}_{ijk})\Pi_{ij}^{\alpha}\times$$
$$
\times(U_{ni}-U_{nj})\biggl[
(2r_{in}^{\mu}-r_{jn}^{\mu})\nabla_{\textbf{R}1}^{\mu}[\Psi^{\dagger}(R',t)\hat{s}_{n}^{\beta}\hat{s}_{j}^{\gamma}\hat{s}_{i}^{\delta}\Psi(R',t)]$$
$$+(2r_{jn}^{\mu}-r_{in}^{\mu})\nabla_{\textbf{R}2}^{\mu}[\Psi^{\dagger}(R',t)\hat{s}_{n}^{\beta}\hat{s}_{j}^{\gamma}\hat{s}_{i}^{\delta}\Psi(R',t)]$$
\begin{equation}\label{MFMUEI P time der I order b}
-(r_{in}^{\mu}+r_{jn}^{\mu})\nabla_{\textbf{R}3}^{\mu}[\Psi^{\dagger}(R',t)\hat{s}_{n}^{\beta}\hat{s}_{j}^{\gamma}\hat{s}_{i}^{\delta}\Psi(R',t)]
\biggr]dR. \end{equation}
Here we use additional short notation $R'\equiv (\textbf{R}_{1}, \textbf{R}_{2}, \textbf{R}_{3}, R_{N-3})$
at $\textbf{R}_{1}= \textbf{R}_{2}= \textbf{R}_{3}= \textbf{R}_{ijk}$.
We also use notation $\nabla_{\textbf{R}1}^{\mu}$ for the space derivative on $\textbf{R}_{1}$, etc.

We can extract the interaction constant from the general expression
(\ref{MFMUEI P time der I order a}) and (\ref{MFMUEI P time der I order b}).
It can be calculated taking integrals over the angles.
So, we can show that this interaction constant is equal to zero.
However, this result is found under assumption that $\Pi_{ij}^{\alpha}$ does not depend on the angles.
Therefore, we consider second part of this expression and show that it is equal to zero anyway.

Considering the correlationless limit as it is described in subsection IV.B
we find the following representation for $\partial_{t}\textbf{P}_{(1a)}$ (\ref{MFMUEI P time der I order a}):
$\partial_{t}\textbf{P}_{(1a)}=g_{1a}^{\alpha\beta}$
$\varepsilon^{\beta\gamma\delta} \nabla^{\mu}(S^{\beta}S^{\gamma}S^{\delta})$
$=0$,
where
$g_{1a}^{\alpha\beta}=(1/3)\int d\textbf{r}_{1}d\textbf{r}_{2}$
$\Pi^{\alpha}(\mid \textbf{r}_{1}-\textbf{r}_{2}\mid)(U(r_{1})-U(r_{2}))$
$(r_{2}^{\beta}-2r_{1}^{\beta})$.

Equation $\partial_{t}\textbf{P}_{(1b)}$ (\ref{MFMUEI P time der I order b})
can be considered as three terms of similar to each other structure.
It would be composed of an interaction constant similar to $g_{1a}^{\alpha\beta}$
and following structure of the spin density vectors $\varepsilon^{\beta\gamma\delta} S^{\beta}S^{\gamma}\nabla^{\mu}S^{\delta}$,
which is equal to zero due to the vector product of two spin densities entering this equation
$\varepsilon^{\beta\gamma\delta} S^{\beta}S^{\gamma}=0$.
Hence we conclude that the polarization $\textbf{P}_{(1)}(\textbf{r},t)=0$
is equal to zero in this order of expansion.

\subsection{Second order expansion}

Let us consider the time derivative of the polarization in the second order on the small interparticle distance.
In the first order expansion
we found that
structures like $(2r_{in}^{\mu}-r_{jn}^{\mu})$, $(2r_{jn}^{\mu}-r_{in}^{\mu})$,
and $(r_{in}^{\mu}+r_{jn}^{\mu})$ appear in front of derivatives
$\nabla_{\textbf{R}1}^{\mu}$, $\nabla_{\textbf{R}2}^{\mu}$, and $\nabla_{\textbf{R}3}^{\mu}$ correspondingly
(and at the expansion of the delta function as well).
Products of pairs of these "bricks"
appears in the second order expansion
(and the products of larger number of these elements would appear in the higher order expansions).
To make our notations shorter we introduce coefficients $C^{\mu\nu}$ as the product of two "bricks":
$C^{\mu\nu}_{11}=(2r_{in}^{\mu}-r_{jn}^{\mu})(2r_{in}^{\nu}-r_{jn}^{\nu})$,
$C^{\mu\nu}_{12}=(2r_{in}^{\mu}-r_{jn}^{\mu})(2r_{jn}^{\nu}-r_{in}^{\nu})$,
$C^{\mu\nu}_{23}=(2r_{jn}^{\mu}-r_{in}^{\mu})(r_{in}^{\nu}+r_{jn}^{\nu})$, etc.
Corresponding part of expansion of equation (\ref{MFMUEI P time der via expl form of comm}) has the following form,
as the sum of three terms of different structure:
$\partial_{t}P^{\alpha}_{(2)}(\textbf{r},t)=$
$\partial_{t}P^{\alpha}_{(2a)}+\partial_{t}P^{\alpha}_{(2b)}+\partial_{t}P^{\alpha}_{(2c)}$.
Here, we present the explicit form of these partial terms.

The first term of this group $\partial_{t}P^{\alpha}_{(2a)}$ is the result of expansion of the delta-function.
Derivatives acting on the delta-function actually act on $\textbf{r}$,
so they can be placed outside of the integral, as we already did it in equation (\ref{MFMUEI P time der I order a})
$$\partial_{t}P^{\alpha}_{(2a)}(\textbf{r},t)=
\frac{1}{2}\frac{1}{3^{2}}\varepsilon^{\beta\gamma\delta}
\partial^{\mu}\partial^{\nu}
\int \sum_{i,j,i\neq j}\sum_{n\neq i,j}\delta(\textbf{r}-\textbf{R}_{ijk})\times$$
\begin{equation}\label{MFMUEI P time der II a order}
\times\Pi_{ij}^{\alpha}(U_{ni}-U_{nj})C^{\mu\nu}_{11}
\Psi^{\dagger}(R',t)\hat{s}_{n}^{\beta}\hat{s}_{j}^{\gamma}\hat{s}_{i}^{\delta}\Psi(R',t)
dR. \end{equation}
In the correlationsless limit
we can rewrite it as
$\partial_{t}P^{\alpha}_{(2a)}\sim \varepsilon^{\beta\gamma\delta}
\partial^{\mu}\partial^{\nu}(S^{\beta}S^{\gamma}S^{\delta})=0$.

The second group $\partial_{t}P^{\alpha}_{(2b)}$ appear as the result of the expansion of the delta-function in the first order
and the expansion of the wave functions up to the first order
$$\partial_{t}P^{\alpha}_{(2b)}=
\frac{1}{3^{2}}\varepsilon^{\beta\gamma\delta}
\partial^{\mu}\int \sum_{i,j,i\neq j}\sum_{n\neq i,j}\delta(\textbf{r}-\textbf{R}_{ijk})\Pi_{ij}^{\alpha}(U_{ni}$$
\begin{equation}\label{MFMUEI P time der II b order}
-U_{nj})
[C^{\mu\nu}_{13}\partial_{R3}^{\nu}-C^{\mu\nu}_{11}\partial_{R1}^{\nu}-C^{\mu\nu}_{12}\partial_{R2}^{\nu}]
\Psi^{\dagger}\hat{s}_{n}^{\beta}\hat{s}_{j}^{\gamma}\hat{s}_{i}^{\delta}\Psi
dR. \end{equation}
Hence, we have one derivative outside of the integral.
Under the integral we have the structure similar to (\ref{MFMUEI P time der I order b}).
Hence, term $\partial_{t}P^{\alpha}_{(2b)}$ is proportional to the derivative of $\partial_{t}P^{\alpha}_{(1b)}$.
It shows that term $\partial_{t}P^{\alpha}_{(2b)}=0$ is equal to zero.
Same conclusion can be seen from the correlationless form of equation (\ref{MFMUEI P time der II b order}).

The last part of $\partial_{t}P^{\alpha}_{(2)}$ appears as the zeroth order expansion of the delta-function
and the expansion of the wave functions up to the second order of expansion.
It has the following structure
$$\partial_{t}P^{\alpha}_{(2c)}=
\frac{1}{2}\frac{1}{3^{2}}\varepsilon^{\beta\gamma\delta}
\int dR\sum_{i,j,i\neq j}\sum_{n\neq i,j}\delta(\textbf{r}-\textbf{R}_{ijk})\Pi_{ij}^{\alpha}(U_{ni}$$ 
$$-U_{nj})[C^{\mu\nu}_{11}\partial_{R1}^{\mu}\partial_{R1}^{\nu}     
+C^{\mu\nu}_{22}\partial_{R2}^{\mu}\partial_{R2}^{\nu}
+C^{\mu\nu}_{33}\partial_{R3}^{\mu}\partial_{R3}^{\nu}
+2C^{\mu\nu}_{12}\partial_{R1}^{\mu}\partial_{R2}^{\nu}$$
\begin{equation}\label{MFMUEI P time der II c order}
-2C^{\mu\nu}_{13}\partial_{R1}^{\mu}\partial_{R3}^{\nu}
-2C^{\mu\nu}_{23}\partial_{R2}^{\mu}\partial_{R3}^{\nu}]
\Psi^{\dagger}\hat{s}_{n}^{\beta}\hat{s}_{j}^{\gamma}\hat{s}_{i}^{\delta}\Psi
. \end{equation}
Equation (\ref{MFMUEI P time der II c order}) gives two groups of terms.
Each of first three terms lead to the following structure
$\varepsilon^{\beta\gamma\delta} (\partial^{\mu}\partial^{\nu}S^{\beta})S^{\gamma}S^{\delta}$.
It is equal to zero due to
$\varepsilon^{\beta\gamma\delta} S^{\gamma}S^{\delta}=0$.
The second group of term is composed of the fourth, fifth and sixth terms.
Each of the has the following structure
$\varepsilon^{\beta\gamma\delta} (\partial^{\mu}S^{\beta})(\partial^{\nu}S^{\gamma})S^{\delta}$.
In general case, it is not equal to zero.
It depends on the symmetry of the interaction constant.
Let us consider the "intermediate" form of the interaction constant appearing from the fourth term:
$\hat{g}_{12}^{\alpha,\mu\nu}=$
$\int d\textbf{r}_{1} d\textbf{r}_{2}(2r_{in}^{\mu}-r_{jn}^{\mu})$
$(2r_{jn}^{\nu}-r_{in}^{\nu})\Pi^{\alpha}(\mid \textbf{r}_{in}-\textbf{r}_{jn}\mid)(U_{ni}-U_{nj})$.
Calculation shows that it is proportional to $\delta^{\mu\nu}$.
It leads to the following modification of the spin density tensor structure
$\varepsilon^{\beta\gamma\delta} (\partial^{\mu}S^{\beta})(\partial^{\mu}S^{\gamma})S^{\delta}=0$.

Therefore, we conclude that all terms in the second order expansion are equal to zero.

In order to consider more general case
we can discuss the contribution of the angle dependence in $\Pi_{ij}^{\alpha}$
and how it contribute in $\partial_{t}P^{\alpha}_{(2c)}$.
Necessary calculations are demonstrated in Appendix A.
It shows the zero contribution of the anisotropy in the second order expansion.

\subsection{Third order expansion}


Previous subsections considering the zeroth, first and second orders of the expansion give the zeroth results even
if we include the anisotropy of the coefficient defining the electric dipole moment of one crystal cell.
The derivation of the polarization evolution in the third order expansion is considered in Appendix B.
Main conclusion is the zero value of the polarization evolution
if we consider isotropic regime for the coefficient defining the electric dipole moment.
The account of the anisotropy partially leads to the nonzero contribution.
We show that
the anisotropic part of function $\Pi^{\alpha}(r,\theta)$ giving nonzero contribution in the third order
is proportional to the first order spherical function $Y_{10}$.
Corresponding part of function $\Pi^{\alpha}(r,\theta)$
enters the interaction constant $g^{\alpha}_{\Pi,l=1}$.

Our calculation gives the following structure for a part of the polarization evolution equation
$\partial_{t}P^{\alpha}_{(3)}=\sum_{i=1}^{6}\partial_{t}P^{\alpha}_{(3)i}$
\begin{equation}\label{MFMUEI P time derivative answer IIIa}
\partial_{t}P^{\alpha}_{(3)i=1}=
-\frac{1}{3!}\frac{\sqrt{3}}{3^{4}}g_{u} g^{\alpha}_{\Pi,l=1}
(\textbf{S}\cdot [\nabla^{z}\textbf{S}\times \triangle\textbf{S}]),
\end{equation}
where we introduced the interaction constant including the anisotropy of the coefficient defining the electric dipole moment
\begin{equation}\label{MFMUEI g Pi l 1 def}
g^{\alpha}_{\Pi,l=1}=
 \int d\textbf{r}' r'\Pi^{\alpha}_{l=1}(r').
\end{equation}
However, the superposition of all terms like (\ref{MFMUEI P time derivative answer IIIa}) gives the zero value
$\partial_{t}P^{\alpha}_{(3)}=0$.

Presence of the angle dependence in function $\Pi^{\alpha}(r,\theta)$ shows existence of the anisotropy axis,
so the z-axis is chosen in the direction of this axis.
This chosen direction is explicitly presented in equation (\ref{MFMUEI P time derivative answer IIIa})
via the derivative of the spin density on the z-projection.

In subsection A we point out
that tensor structure similar to (\ref{MFMUEI P time derivative answer IIIa}) can be expected
from the qualitative analysis of the polarization definition and the form of the Hamiltonian.
Moreover, one of the derivatives in this structure has fixed projection on the $z$-direction,
while the Laplacian $\triangle\textbf{S}$ is formed out of two other space derivatives.

The role of the anisotropic part of function $\Pi^{\alpha}(r,\theta)$ can be estimated
from its static value (it is discussed below) and the comparison of the presented model
with the experiment.
In the next subsection we show
that the isotropic part of function $\Pi^{\alpha}(r,\theta)$ gives major contribution.
Hence, we need to continue our derivation and consider next order of the expansion.

\subsection{Fourth order expansion}

Let us repeat that we are mostly focused on the regime of the isotropic constant $\Pi^{\alpha}(r)$
defining the electric dipole moment of the crystal cell due to the symmetric exchange interaction
(the exchange part of the Coulomb interaction).
In this regime we have zero value of the interaction in the lowest order as
we described it above.
The fourth order on the small interparticle distance gives nonzero contribution.
Some details of the calculations are demonstrated in Appendix C.
Let us repeat the result derived in Appendix C:
\begin{equation}\label{MFMUEI P time derivative answer IV}
\partial_{t}P^{\alpha}_{(4)}
=G(\textbf{S}\cdot [\nabla^{\mu}\textbf{S}\times \nabla^{\mu}\triangle\textbf{S}]), \end{equation}
where
\begin{equation}\label{MFMUEI G1 main}
G=\frac{1}{3!}\frac{1}{3}g_{u}g_{\Pi}^{\alpha},
\end{equation}
is a combined interaction constant
with
$g_{u}=\int r^{2}U(r)d^{3}r$,
and
$g_{\Pi}^{\alpha}=\int r^{2}\Pi^{\alpha}(r)d^{3}r$.

Equation (\ref{MFMUEI P time derivative answer IV}) is the main result of this paper
which presents the polarization evolution equation in the exchange-strictionally formed II-type multiferroics.

Interaction appears in equation (\ref{MFMUEI P time derivative answer IV}) in highly nonlinear form.
So, there is no contribution of the this interaction in the small amplitude perturbations of the spin density at the uniform equilibrium.
However, it can be nonzero for the periodical equilibrium structures including spin spirals and their perturbations.

The right-hand side of equation (\ref{MFMUEI P time derivative answer IV}) can be presented as the divergence of the second rank tensor
$J_{P}^{\alpha\mu}=-G(\textbf{S}\cdot [\nabla^{\mu}\textbf{S}\times \triangle\textbf{S}])$.
Here, function $J_{P}^{\alpha\mu}$ is an effective polarization current.
Hence, equation (\ref{MFMUEI P time derivative answer IV}) can be considered as an effective continuity equation
$\partial_{t}P^{\alpha}+\nabla^{\mu}J_{P}^{\alpha\mu}=0$
showing the conservation of each projection of the polarization.
The spin evolution equation can be also considered as an effective continuity equation in this regime.
So, equation (\ref{MFMUEI LL eq dissipless}) can be represented as
$\partial_{t}S^{\alpha}+\nabla^{\mu}J_{S}^{\alpha\mu}=0$,
where
$J_{S}^{\alpha\mu}=-g_{u}\varepsilon^{\alpha\beta\gamma}S^{\beta}\nabla^{\mu}S^{\gamma}$
is the effective spin current.

\subsection{Structure of the polarization definition}

Equation of the polarization evolution is obtained in the form of expansion of the wave function
under the integral after replacement of the time derivative by the Hamiltonian of the system.
In order to interpret found equation (\ref{MFMUEI P time derivative answer IV})
we need to understand the possibility of presence of the polarization on the right-hand side of equation (\ref{MFMUEI P time derivative answer IV}).
We need to check that
the polarization does not enter the right-hand side of equation (\ref{MFMUEI P time derivative answer IV}).
Hence, it can be left in the obtained form as the function of the spin densities.
To answer this question we need to expand the definition of polarization
(\ref{MFMUEI P def})
on the relative distance
and check its contribution on the right-hand side of equation (\ref{MFMUEI P time derivative answer IV}).
Before we present calculations we can point out the structure of the spin operators in the polarization definition,
where we see the scalar product of the spin operators.
In contrast to it see the vector product of the spin densities in the polarization evolution equation caused by the exchange interaction.

The polarization definition
(\ref{MFMUEI P def})
contains one undefinite short-range function $\Pi^{\alpha}(\textbf{r})$.
We explicitly include the dynamic of two particles with numbers $i$ and $j$.
It happens in contrast with the equation for the polarization evolution,
where we consider three particles.
Therefore, we introduce the center of mass and coordinate of relative motion for two particles
and present the expansion on the small relative distance.
Otherwise, function $\Pi^{\alpha}(\textbf{r})$ is equal to zero, at the larger distances.
If we consider regime of isotropic function $\Pi^{\alpha}(r)$
we have nonzero contribution in the zeroth and second order of expansion.
It gives us major term and a correction to it.
Therefore, we do not consider further expansion.
So, we find the following result for the approximate form of the polarization
\begin{equation}\label{MFMUEI P appr iso}
P^{\alpha}_{A}=
g_{0\Pi}^{\alpha} \textbf{S}^{2}(\textbf{r},t)
+\frac{1}{3}\frac{1}{2^{2}}g_{\Pi}^{\alpha}[\triangle \textbf{S}^{2} -2 (\partial_{\mu} S^{\nu})(\partial_{\mu} S^{\nu})],
\end{equation}
where subindex $A$ refers to the approximate form of the polarization.

To give some analysis of equation (\ref{MFMUEI P appr iso}) let us to calculate the polarization for two spin structures.
First, we consider the uniform system of parallel spins $\textbf{S}=S_{0}\textbf{e}_{z}$,
where $S_{0}$ is the constant.
The constant value of the polarization appears from the first term $P^{\alpha}_{A}=
g_{0\Pi}^{\alpha}S_{0}^2$.
Next, we consider the cycloidal spiral spin structure
$\textbf{S}_{0}(y)=s_{b}\textbf{e}_{y}\cos(yq)
+s_{c}\textbf{e}_{z}\sin(yq)
+s_{a}\textbf{e}_{x}$.
Both terms in equation (\ref{MFMUEI P appr iso}) gives nonzero contributions.
Consider a regime with $\mid s_{b}\mid=\mid s_{c}\mid$.
It leads to superposition of two constant contribution from both terms in equation (\ref{MFMUEI P appr iso}):
$P^{\alpha}_{A}=
g_{0\Pi}^{\alpha}(s_{b}^{2}+s_{a}^{2})
-\frac{1}{6}g_{\Pi}^{\alpha}q^{2}s_{b}^{2}S$.
If $\mid s_{b}\mid\neq \mid s_{c}\mid$
we see that polarization is a periodic function similarly to the spin:
$P^{\alpha}_{A}=
g_{0\Pi}^{\alpha}[s_{a}^{2}+(s_{b}^{2}+s_{c}^{2})/2+(1/2)\cos(2qy)(s_{b}^{2}-s_{c}^{2})]
-\frac{1}{12}g_{\Pi}^{\alpha}q^{2}[s_{b}^{2}+s_{c}^{2}+\cos(2qy)(s_{b}^{2}-s_{c}^{2})]$,
but it shows the periodic change of the polarization with the double wave vector.

If we include the anisotropy of function
we find additional contribution appearing in the first order on the relative distance
\begin{equation}\label{MFMUEI P appr ani}
P^{\alpha}_{A}=
-\frac{1}{2\sqrt{3}}\biggl(\int d\textbf{r}' r' \Pi_{l=1}^{\alpha}(r')\biggr)
\partial_{z}\textbf{S}^{2} .\end{equation}

These approximate expressions show that the polarization evolution does not required further representation via the polarization
and we can apply the expression in terms of the spin density.

\subsection{On a simplified derivation of the polarization evolution equation}

The previous subsection presents approximate relations between polarization (\ref{MFMUEI P def})
and the spin density (\ref{MFMUEI S def}).
Moreover, we have equation for the spin evolution equation (\ref{MFMUEI LL eq dissipless}).
However, we need to point out that at the derivation of the approximate
polarization (\ref{MFMUEI P appr iso}) and (\ref{MFMUEI P appr ani})
we used expansion on the two-particle relative distance.
Same approximate structure is used for the derivation of spin evolution equation (\ref{MFMUEI LL eq dissipless}).
So, the direct relation to the third particle is broken here.
Can we use it in order to derive the polarization evolution equation?
Let us present the result for each part of the polarization.

In the zeroth order of the expansion we get the time derivative of the scalar product of the spin densities.
It leads us to the zero value of the time derivative.
It happens since the time derivative is proportional to the vector product containing the spin density
\begin{equation}\label{MFMUEI P A der 0}
\partial_{t}P^{\alpha}_{A,0}
=2g_{0\Pi}^{\alpha} \textbf{S}\partial_{t}\textbf{S}
=\frac{1}{3}g_{0\Pi}^{\alpha}g_{u}\varepsilon^{\beta\gamma\delta}S^{\beta}S^{\gamma}\triangle S^{\delta}=0 ,
\end{equation}
where index $0$ refers to the analysis of the polarization appearing in the zeroth order of the expansion.

In the second order of the expansion we have two terms in equation (\ref{MFMUEI P appr iso}).
The first term is proportional to $\textbf{S}^{2}$.
It gives the zero contribution in the polarization evolution.
So, we find the contribution of the second term only.
It leads to the following expression
\begin{equation}\label{MFMUEI P A der 2}
\partial_{t}P^{\alpha}_{A,2}=\frac{1}{18} g_{\Pi}^{\alpha}g_{u}
\varepsilon^{\beta\gamma\delta}S^{\beta}(\partial_{\mu}S^{\gamma})\partial_{\mu}\triangle S^{\delta}.
\end{equation}
We see that the simplified derivation based on the account of the second term in the expansion (\ref{MFMUEI P appr iso})
gives equation (\ref{MFMUEI P A der 2}) coinciding with the equation obtained via the complete derivation (\ref{MFMUEI P time derivative answer IV}).

\section{Conclusion}

The method of quantum hydrodynamics has been developed for the multiferroics.
This method originally developed for the quantum plasmas-like mediums of charged particles
and quantum gases of neutral atoms, including the spinor Bose-Einstein condensates.
Its development for the multiferroics follows the work of the application of quantum hydrodynamic method to major features of the ferromagnetic materials.
All together, this work shows wide range of the applicability of the quantum hydrodynamics.
It also extends its applicability range to the multiferroics.
Definitely, we have considered one feature of the multiferroics.
We have made derivation of the electric polarization evolution equation for II-type multiferroics
with the electrostriction mechanism of the electric dipole formation,
where we have considered the evolution under the Coulomb exchange interaction given by the Heisenberg Hamiltonian.
However, it shows a path for the account of other interactions in the evolution of polarization in all types of multiferroics.
One mechanism of the polarization evolution has been considered,
it is the Coulomb exchange interaction.
It appears in the fourth order expansion on the small interparticle distance.
It gives the following tensor structure of the spin densities
$\partial_{t}P^{\alpha}
\sim\varepsilon^{\beta\gamma\delta}
S^{\beta}\cdot \partial^{\mu}S^{\gamma}\cdot\partial^{\mu}\triangle S^{\delta}$,
with the coefficient depending on the microscopic parameters.

\section{Acknowledgements}

The work of M.I. Trukhanova is supported by the Russian Science Foundation under the
grant No. 22-72-00036.

\section{DATA AVAILABILITY}

Data sharing is not applicable to this article as no new data were
created or analyzed in this study, which is a purely theoretical one.

\section{Author Contribution Statement}

All authors contributed equally to the paper.

\appendix

\section{Anisotropy contribution in the second order expansion}

We need to consider the contribution of the anisotropy of function $\Pi_{ij}^{\alpha}$
in the last three terms of equation (\ref{MFMUEI P time der II c order}).
Corresponding terms can be presented in the following form
$\partial_{t}P^{\alpha}_{(2c)}=
\frac{1}{3^{2}}\varepsilon^{\beta\gamma\delta}
S^{\beta}\cdot \partial^{\nu}S^{\gamma}\cdot\partial^{\mu} S^{\delta}
[g_{12}^{\alpha,\mu\nu}+g_{13}^{\alpha,\mu\nu}-g_{23}^{\alpha,\mu\nu}]$,
where
$g_{ab}^{\alpha,\mu\nu}=\int d\textbf{r}_{1}d\textbf{r}_{2} C_{ab}^{\mu\nu}(\textbf{r}_{1},\textbf{r}_{2}) \Pi^{\alpha}(\textbf{r}_{1}-\textbf{r}_{2})[U(r_{1})-U(r_{2})]$.
So, the tensor constant $g_{ab}^{\alpha,\mu\nu}$ can be presented as the superposition of two constants:
$g_{ab}^{\alpha,\mu\nu}=R_{1ab}^{\alpha,\mu\nu}-R_{2ab}^{\alpha,\mu\nu}$,
where
$R_{1ab}^{\alpha,\mu\nu}=\int d\textbf{r}_{1}d\textbf{r}_{2} C_{ab}^{\mu\nu}(\textbf{r}_{1},\textbf{r}_{2}) \Pi^{\alpha}(\textbf{r}_{1}-\textbf{r}_{2})U(r_{1})$,
with the further transition to coordinates $\textbf{r}_{1}$ and $\textbf{r}_{3}=\textbf{r}_{1}-\textbf{r}_{2}$.
Hence, we have
$R_{1ab}^{\alpha,\mu\nu}=\int d\textbf{r}_{1}U(r_{1})d\textbf{r}_{3} C_{ab}^{\mu\nu}(\textbf{r}_{1},\textbf{r}_{3}) \Pi^{\alpha}(\textbf{r}_{3})$.
Here we consider $\Pi^{\alpha}(\textbf{r}_{3})$ instead of $\Pi^{\alpha}(r_{3})$ to include the anisotropy of this function.
Parameter $R_{1ab}^{\alpha,\mu\nu}$ appears as a superposition of products of integrals.
It includes three types of combinations of the integrals.
First combination is
$\int d\textbf{r}_{1} r_{1}^{\mu}U(r_{1})\cdot\int d\textbf{r}_{3}r_{3}^{\nu} \Pi^{\alpha}(\textbf{r}_{3})$,
where $\int d\textbf{r}_{1} r_{1}^{\mu}U(r_{1})=0$.
So, the anisotropy of $\Pi^{\alpha}(\textbf{r}_{3})$ does not change this kinds of terms.

Let us to point out that we consider the rotational symmetry and include the dependence on angle $\theta$ only.
$\Pi^{\alpha}(\textbf{r}_{3})=\Pi^{\alpha}(r_{3},\theta_{3})$.
We consider it via the expansion on the spherical functions $Y_{l0}(\cos\theta)$:
$\Pi^{\alpha}(r_{3},\theta_{3})=\sqrt{4\pi}\sum_{l=0}^{\infty}\Pi_{l}^{\alpha}(r_{3})Y_{l0}(\cos\theta)$,
with the normalization $\Pi_{l=0}^{\alpha}(r_{3})=\Pi^{\alpha}(r_{3})$.

Second combination is
$\int d\textbf{r}_{1} r_{1}^{\mu}r_{1}^{\nu}U(\textbf{r}_{1})\cdot\int d\textbf{r}_{3} \Pi^{\alpha}(\textbf{r}_{3})$.
Integral $\int d\textbf{r}_{3} \Pi^{\alpha}(\textbf{r}_{3})$ gives the nonzero contribution of $\Pi_{l=0}^{\alpha}(r_{3})=\Pi^{\alpha}(r_{3})$ only.

Third combination is
$\int d\textbf{r}_{1}U(\textbf{r}_{1})\cdot\int d\textbf{r}_{3}r_{3}^{\mu}r_{3}^{\nu} \Pi^{\alpha}(\textbf{r}_{3})$.
Hence, we need to consider integral $\hat{I}^{\mu\nu}\equiv \int d\textbf{r}_{3}r_{3}^{\mu}r_{3}^{\nu} \Pi^{\alpha}(r_{3},\theta_{3})$
(the matrix composed of integrals).
We do not consider dependence of $\Pi^{\alpha}$ on $\varphi$.
Consequently, this integral in diagonal
$$\hat{I}= \int d\textbf{r}_{3} \Pi^{\alpha}(r_{3},\theta_{3})\left(
                                                               \begin{array}{ccc}
                                                                 \frac{1}{2}\sin^{2}\theta_{3} & 0 & 0 \\
                                                                 0 & \frac{1}{2}\sin^{2}\theta_{3} & 0 \\
                                                                 0 & 0 & \cos^{2}\theta_{3} \\
                                                               \end{array}
                                                             \right),
$$
with
$\frac{1}{2}\sin^{2}\theta_{3}=(\sqrt{4\pi}/3)Y_{00}-(2\sqrt{\pi}/3\sqrt{5})Y_{20}$,
and
$\cos^{2}\theta_{3}=(\sqrt{4\pi}/3)Y_{00} +(4\sqrt{\pi}/3\sqrt{5})Y_{20}$,
where $Y_{00}=1/\sqrt{4\pi}$, and $Y_{20}=\sqrt{5/16\pi}(3\cos^{2}\theta-1)$.

Calculation gives the following result
$\hat{I}^{\mu\nu}=\frac{1}{3}(\delta^{\mu\nu}\int d\textbf{r}_{3} \Pi_{l=0}^{\alpha}(r_{3})
+I_{2}^{\mu\nu}\frac{1}{\sqrt{5}}\int d\textbf{r}_{3} \Pi_{l=2}^{\alpha}(r_{3}))$,
with
$$I_{2}\equiv\left(
                                                               \begin{array}{ccc}
                                                                 -1 & 0 & 0 \\
                                                                 0 & -1 & 0 \\
                                                                 0 & 0 & 2 \\
                                                               \end{array}
                                                             \right).$$
Product of $\varepsilon^{\beta\gamma\delta}
S^{\beta}\cdot \partial^{\nu}S^{\gamma}\cdot\partial^{\mu} S^{\delta}$ and $\delta^{\mu\nu}$ gives zero value,
but product of $\varepsilon^{\beta\gamma\delta}
S^{\beta}\cdot \partial^{\nu}S^{\gamma}\cdot\partial^{\mu} S^{\delta}$ and $I_{2}^{\mu\nu}$ gives
$\varepsilon^{\beta\gamma\delta}
(-S^{\beta}\cdot \partial^{x}S^{\gamma}\cdot\partial^{x} S^{\delta}
-S^{\beta}\cdot \partial^{y}S^{\gamma}\cdot\partial^{y} S^{\delta}
+2S^{\beta}\cdot \partial^{z}S^{\gamma}\cdot\partial^{z} S^{\delta})=0$,
which is equal to zero as well.
We obtain same results for other term and give conclusion $\partial_{t}P^{\alpha}_{(2c)}=0$ even for the anisotropic $\Pi_{ij}^{\alpha}$.

\section{Polarization in the third order expansion}

We present the time derivative of the polarization in the third order on the interparticle distance
$\partial_{t}P^{\alpha}_{(3)}$ as the sum of four terms:
$\partial_{t}P^{\alpha}_{(3)}=\partial_{t}P^{\alpha}_{(3a)}$
$+\partial_{t}P^{\alpha}_{(3b)}+\partial_{t}P^{\alpha}_{(3c)}$
$+\partial_{t}P^{\alpha}_{(3d)}$.
As the first part of $\partial_{t}P^{\alpha}_{(3)}$
we consider the term appearing from the expansion of the delta-function in the third order and expansion of the wave functions in the zeroth order
$$\partial_{t}P^{\alpha}_{(3a)}=
-\frac{1}{3!}\frac{1}{3^{3}}\varepsilon^{\beta\gamma\delta}
\partial^{\mu}\partial^{\nu}\partial^{\sigma}\int \sum_{i,j,i\neq j}\sum_{n\neq i,j}\delta(\textbf{r}-\textbf{R}_{ijk})\times$$
\begin{equation}\label{MFMUEI P time der III order a}
\times\Pi_{ij}^{\alpha}(U_{ni}-U_{nj})C^{\mu\nu\sigma}_{111}\Psi^{\dagger}(R',t)\hat{s}_{n}^{\beta}\hat{s}_{j}^{\gamma}\hat{s}_{i}^{\delta}
\Psi(R',t)dR, \end{equation}
where
the third rank tensor $C^{\mu\nu\sigma}$ composed of the coefficients of the expansion has the following structure:
$C^{\mu\nu\sigma}_{112}=(2r_{in}^{\mu}-r_{jn}^{\mu})(2r_{in}^{\nu}-r_{jn}^{\nu})(2r_{jn}^{\sigma}-r_{in}^{\sigma})$,
$C^{\mu\nu\sigma}_{123}=(2r_{in}^{\mu}-r_{jn}^{\mu})(2r_{jn}^{\nu}-r_{in}^{\nu})(r_{in}^{\sigma}+r_{jn}^{\sigma})$, etc.
It is proportional to the third space derivative of the term appearing in the zeroth order of the expansion
(\ref{MFMUEI P time der 0 order fin}).
Hence, it is equal to zero $\partial_{t}P^{\alpha}_{(3a)}=0$.

As the second term we choose the term containing the expansion of the delta-function in the second order
and the expansion of the wave functions up to the first order:
$$\partial_{t}P^{\alpha}_{(3b)}=
\frac{1}{2!}\frac{1}{3^{3}}\varepsilon^{\beta\gamma\delta}
\partial^{\mu}\partial^{\nu}\int \sum_{i,j,i\neq j}\sum_{n\neq i,j}\delta(\textbf{r}-\textbf{R}_{ijk})\Pi_{ij}^{\alpha}\times$$
\begin{equation}\label{MFMUEI P time der III order b}
\times(U_{ni}-U_{nj})
[C^{\mu\nu\sigma}_{111}\partial_{R1}^{\sigma}+C^{\mu\nu\sigma}_{112}\partial_{R2}^{\sigma}
-C^{\mu\nu\sigma}_{113}\partial_{R3}^{\sigma}]
\Psi^{\dagger}\hat{s}_{n}^{\beta}\hat{s}_{j}^{\gamma}\hat{s}_{i}^{\delta}
\Psi dR. \end{equation}
It appears as the superposition of terms proportional to
$\varepsilon^{\beta\gamma\delta}
\partial^{\mu}\partial^{\nu}
[(\partial^{\sigma}S^{\beta})S^{\gamma}S^{\delta}]=0$.
Hence, we obtain $\partial_{t}P^{\alpha}_{(3b)}=0$.

Next, we consider the term appearing at the expansion of the delta-function in the first order
and expansion of the wave functions up to the second order:
$$\partial_{t}P^{\alpha}_{(3c)}=
-\frac{1}{2!}\frac{1}{3^{3}}\varepsilon^{\beta\gamma\delta}
\partial^{\mu}\int \sum_{i,j,i\neq j}\sum_{n\neq i,j}\delta(\textbf{r}-\textbf{R}_{ijk})\times$$
$$\times\Pi_{ij}^{\alpha}(U_{ni}-U_{nj})
[C^{\mu\nu\sigma}_{111}\partial_{R1}^{\nu}\partial_{R1}^{\sigma}     
+C^{\mu\nu\sigma}_{122}\partial_{R2}^{\nu}\partial_{R2}^{\sigma}$$
$$+C^{\mu\nu\sigma}_{133}\partial_{R3}^{\nu}\partial_{R3}^{\sigma}
+2C^{\mu\nu\sigma}_{112}\partial_{R1}^{\nu}\partial_{R2}^{\sigma}$$
\begin{equation}\label{MFMUEI P time der III order c}
-2C^{\mu\nu\sigma}_{113}\partial_{R1}^{\nu}\partial_{R3}^{\sigma}
-2C^{\mu\nu\sigma}_{123}\partial_{R2}^{\nu}\partial_{R3}^{\sigma}]
\Psi^{\dagger}\hat{s}_{n}^{\beta}\hat{s}_{j}^{\gamma}\hat{s}_{i}^{\delta}
\Psi dR. \end{equation}
Each of first three terms in this equation is proportional to
$\varepsilon^{\beta\gamma\delta}
\partial^{\mu}[(\partial^{\nu}\partial^{\sigma}S^{\beta})S^{\gamma}S^{\delta}]=0$.
So, they are equal to zero.
Three other terms are proportional to structure like
$\varepsilon^{\beta\gamma\delta}
\partial^{\mu}[(\partial^{\nu}S^{\beta})(\partial^{\sigma}S^{\gamma})S^{\delta}]$.
It can be equal to zero if there is symmetry on indexes $\nu$ and $\sigma$.
Hence, we need to consider the coefficient in front of this term,
which appears as the integral of functions $U(r)$, $\Pi^{\alpha}(\textbf{r})$, and coefficients $C^{\mu\nu\sigma}_{abc}$.
In the isotropic regime $\Pi^{\alpha}(\textbf{r})=\Pi^{\alpha}(r)$,
these integrals are equal to zero, so their tensor structure is not essential.
If we include the anisotropy of function $\Pi^{\alpha}(\textbf{r})=\Pi^{\alpha}(r,\theta)$,
we find three nonzero terms, but each of them is symmetric on indexes $\nu$ and $\sigma$.
Therefore, we obtain
$\partial_{t}P^{\alpha}_{(3c)}=0$.

The last part of $\partial_{t}P^{\alpha}_{(3)}$ is presented as $\partial_{t}P^{\alpha}_{(3d)}$.
It is obtained at the expansion of the delta-function in the zeroth order and the expansion of the wave functions up to the third order:
$$\partial_{t}P^{\alpha}_{(3d)}=
\frac{1}{3!}\frac{1}{3^{3}}\varepsilon^{\beta\gamma\delta}
\int \sum_{i,j,i\neq j}\sum_{n\neq i,j}\delta(\textbf{r}-\textbf{R}_{ijk})\times$$
$$\times\Pi_{ij}^{\alpha}(U_{ni}-U_{nj})
[C^{\mu\nu\sigma}_{111}\partial_{R1}^{\mu}\partial_{R1}^{\nu}\partial_{R1}^{\sigma}     
+C^{\mu\nu\sigma}_{222}\partial_{R2}^{\mu}\partial_{R2}^{\nu}\partial_{R2}^{\sigma}$$
$$-C^{\mu\nu\sigma}_{333}\partial_{R3}^{\mu}\partial_{R3}^{\nu}\partial_{R3}^{\sigma}
+3C^{\mu\nu\sigma}_{112}\partial_{R1}^{\mu}\partial_{R1}^{\nu}\partial_{R2}^{\sigma}
-3C^{\mu\nu\sigma}_{113}\partial_{R1}^{\mu}\partial_{R1}^{\nu}\partial_{R3}^{\sigma}$$
$$+3C^{\mu\nu\sigma}_{122}\partial_{R1}^{\mu}\partial_{R2}^{\nu}\partial_{R2}^{\sigma}
+3C^{\mu\nu\sigma}_{133}\partial_{R1}^{\mu}\partial_{R3}^{\nu}\partial_{R3}^{\sigma}
-3C^{\mu\nu\sigma}_{223}\partial_{R2}^{\mu}\partial_{R2}^{\nu}\partial_{R3}^{\sigma}$$

\begin{equation}\label{MFMUEI P time der III order d}
+3C^{\mu\nu\sigma}_{233}\partial_{R2}^{\mu}\partial_{R3}^{\nu}\partial_{R3}^{\sigma}
-6C^{\mu\nu\sigma}_{123}\partial_{R1}^{\mu}\partial_{R2}^{\nu}\partial_{R3}^{\sigma}]
\Psi^{\dagger}\hat{s}_{n}^{\beta}\hat{s}_{j}^{\gamma}\hat{s}_{i}^{\delta}
\Psi dR. \end{equation}
Function $\partial_{t}P^{\alpha}_{(3d)}$ can be considered as a superposition of three group of terms.
The first group is composed of the first three terms.
Let us consider the spin structure of these terms in the correlationless limit.
For the first term we have the following structure
$\varepsilon^{\beta\gamma\delta}
S^{\beta}S^{\gamma}\partial^{\mu}\partial^{\nu}\partial^{\sigma}S^{\delta}=0$
due to the symmetry of the spins on $\beta$ and $\gamma$.
Same conclusion is correct for two other terms.

The second group of terms is composed of six terms with numbers from number four to the number nine.
All of them are proportional to the number $3$.
Let us introduce notation for this group of terms $\partial_{t}P^{\alpha}_{(3d,II)}$.
All these terms are proportional to
$\varepsilon^{\beta\gamma\delta}
S^{\beta}(\partial^{\sigma}S^{\gamma})\partial^{\mu}\partial^{\nu}S^{\delta}\neq0$
and this tensor structure does not equal to zero.
So, we need to consider the coefficient composed of integrals of functions $U(r)$ and $\Pi^{\alpha}(\textbf{r})$.
First, we consider isotropic limit for $\Pi^{\alpha}(\textbf{r})$.
The coefficient is the superposition of six integrals of the following structure
$g_{abc}^{\alpha,\mu\nu\sigma}=\int d\textbf{r}_{1}d\textbf{r}_{2} C_{abc}^{\mu\nu\sigma}(\textbf{r}_{1},\textbf{r}_{2}) \Pi^{\alpha}(\mid\textbf{r}_{1}-\textbf{r}_{2}\mid)[U(r_{1})-U(r_{2})]$.
We can decompose it on two terms $g_{abc}^{\alpha,\mu\nu\sigma}=R_{1,abc}^{\alpha,\mu\nu\sigma}-R_{2,abc}^{\alpha,\mu\nu\sigma}$
related to $U(r_{1})$ and $U(r_{2})$ correspondingly.
In function $R_{1,abc}^{\alpha,\mu\nu\sigma}$
(in function $R_{2,abc}^{\alpha,\mu\nu\sigma}$)
we make transition to coordinates $\textbf{r}_{1}$ and $\textbf{r}_{3}$
(to coordinates $\textbf{r}_{2}$ and $\textbf{r}_{3}$).
Hence, for instance, $R_{1,abc}^{\alpha,\mu\nu\sigma}$ appears as the superposition of products of two integrals like:
$(\int d\textbf{r} U(r))$$(\int d\textbf{r} r^{\mu}r^{\nu}r^{\sigma}\Pi^{\alpha}(r))$,
$(\int d\textbf{r} r^{\sigma}U(r))$$(\int d\textbf{r} r^{\mu}r^{\nu}\Pi^{\alpha}(r))$,
$(\int d\textbf{r} r^{\nu}r^{\sigma}U(r))$$(\int d\textbf{r} r^{\mu}\Pi^{\alpha}(r))$,
and
$(\int d\textbf{r} r^{\mu}r^{\nu}r^{\sigma}U(r))$$(\int d\textbf{r}\Pi^{\alpha}(r))$.
Each of these four terms contains integral, which is equal to zero:
$(\int d\textbf{r} r^{\mu}\Pi^{\alpha}(r))=0$,
$(\int d\textbf{r} r^{\mu}U(r))=0$,
$(\int d\textbf{r} r^{\mu}r^{\nu}r^{\sigma}U(r))=0$, etc.
Consequently, the group of terms under consideration is equal to zero in the limit of isotropic function $\Pi^{\alpha}$.

Let us consider the anisotropy of function $\Pi^{\alpha}(\textbf{r})$ in the second group of terms.
General structure of the spin densities is the same as in the isotropic regime demonstrated above.
All difference in the calculation of integrals in $g_{abc}^{\alpha,\mu\nu\sigma}$.
Integrals of the interaction potential $U(r)$ with uneven number of the projection of coordinates are equal to zero
$(\int d\textbf{r} r^{\mu}U(r))=0$, $(\int d\textbf{r} r^{\mu}r^{\nu}r^{\sigma}U(r))=0$.
However, similar integrals with function $\Pi^{\alpha}(r,\theta)$ should be calculated
after expansion of function $\Pi^{\alpha}(r,\theta)$ on the spherical functions $Y_{l,0}=Y(\cos\theta)$.
Here we need to calculate the following combination
$g_{112}+g_{113}-g_{221}+g_{331}-g_{223}-g_{332}$,
which is equal to zero due to the combination of the nonzero elements.
As the result of presented analysis
we find
$\partial_{t}P^{\alpha}_{(3d,II)}=0$.

The last group of terms composed of single term,
which is the last term in equation (\ref{MFMUEI P time der III order d}).
It can be presented in the following form
$\partial_{t}P^{\alpha}_{(3d,III)}=-6(1/3!)(1/3^{3})\varepsilon^{\beta\gamma\delta} g^{\alpha,\mu\nu\sigma}_{123}
(\partial^{\sigma}S^{\beta})(\partial^{\nu}S^{\gamma})\partial^{\mu}S^{\delta}$.
Our calculation shows that it is equal to zero even if we include the anisotropy of function $\Pi^{\alpha}(\textbf{r})$.

Regarding $\partial_{t}P^{\alpha}_{(3)}$
we can conclude that it is equal to zero.

\section{Polarization in the fourth order expansion}

It can be presented as a combination of five terms:
$\partial_{t}P^{\alpha}_{(4)}=\partial_{t}P^{\alpha}_{(4a)}$
$+\partial_{t}P^{\alpha}_{(4b)}+\partial_{t}P^{\alpha}_{(4c)}$
$+\partial_{t}P^{\alpha}_{(4d)}+\partial_{t}P^{\alpha}_{(4e)}$.
Let us present each of them.

The smallest group of terms in $\partial_{t}P^{\alpha}_{(4)}$ appears
if we consider the expansion of the delta-function in the fourth order on the interparticle distance
and combine it with the expansion of the wave functions in the zeroth order:
$$\partial_{t}P^{\alpha}_{(4a)}=
\frac{1}{4!}\frac{1}{3^{4}}\varepsilon^{\beta\gamma\delta}
\partial^{\mu}\partial^{\nu}\partial^{\sigma}\partial^{\lambda}\int \sum_{i,j,i\neq j}\sum_{n\neq i,j}\delta(\textbf{r}-\textbf{R}_{ijk})\times$$
\begin{equation}\label{MFMUEI P time der IV order a}
\times\Pi_{ij}^{\alpha}(U_{ni}-U_{nj})
C^{\mu\nu\sigma\lambda}_{1111}\Psi^{\dagger}\hat{s}_{n}^{\beta}\hat{s}_{j}^{\gamma}\hat{s}_{i}^{\delta}
\Psi dR, \end{equation}
where
the fourth rank tensor $C^{\mu\nu\sigma\lambda}$ composed of the coefficients of the expansion has the following structure:
$C^{\mu\nu\sigma\lambda}_{1122} =(2r_{in}^{\mu}-r_{jn}^{\mu})$
$(2r_{in}^{\nu}-r_{jn}^{\nu})(2r_{jn}^{\sigma}-r_{in}^{\sigma})$
$(2r_{jn}^{\lambda}-r_{in}^{\lambda})$,
$C^{\mu\nu\sigma\lambda}_{1233}=(2r_{in}^{\mu}-r_{jn}^{\mu})$
$(2r_{jn}^{\nu}-r_{in}^{\nu})(r_{in}^{\sigma}+r_{jn}^{\sigma})$
$(r_{in}^{\lambda}+r_{jn}^{\lambda})$, etc.
It is proportional to the fourth space derivative of the term appearing in the zeroth order of the expansion
(\ref{MFMUEI P time der 0 order fin}).
Hence, it is equal to zero $\partial_{t}P^{\alpha}_{(4a)}=0$.

As the second term we choose the term containing the expansion of the delta-function in the third order
and the expansion of the wave functions up to the first order:
$$\partial_{t}P^{\alpha}_{(4b)}=
-\frac{1}{3!}\frac{1}{3^{4}}\varepsilon^{\beta\gamma\delta}
\partial^{\mu}\partial^{\nu}\partial^{\sigma}\int \sum_{i,j,i\neq j}\sum_{n\neq i,j}\delta(\textbf{r}-\textbf{R}_{ijk})\times$$
$$\times\Pi_{ij}^{\alpha}(U_{ni}-U_{nj})
[C^{\mu\nu\sigma\lambda}_{1111}\partial_{R1}^{\lambda}$$
\begin{equation}\label{MFMUEI P time der IV order b}
+C^{\mu\nu\sigma\lambda}_{1112}\partial_{R2}^{\lambda}
-C^{\mu\nu\sigma\lambda}_{1113}\partial_{R3}^{\lambda}]
\Psi^{\dagger}\hat{s}_{n}^{\beta}\hat{s}_{j}^{\gamma}\hat{s}_{i}^{\delta}
\Psi dR. \end{equation}
It appears as the superposition of terms proportional to
$\varepsilon^{\beta\gamma\delta}
\partial^{\mu}\partial^{\nu}\partial^{\sigma}
((\partial^{\lambda}S^{\beta})S^{\gamma}S^{\delta})=0$.
Hence, we obtain $\partial_{t}P^{\alpha}_{(4b)}=0$.

The third term in this sequence is the term appearing at the expansion of the delta-function in the second order on the interparticle distance.
Consequently, we need to consider the expansion of two wave functions up to the second order of the expansion, so their product gives the second order on the interparticle distance.
The result of the described expansion has the following structure:
$$\partial_{t}P^{\alpha}_{(4c)}=
\frac{1}{2!}\frac{1}{2!} \frac{1}{3^{4}}\varepsilon^{\beta\gamma\delta}
\partial^{\mu}\partial^{\nu}\int \sum_{i,j,i\neq j}\sum_{n\neq i,j}\delta(\textbf{r}-\textbf{R}_{ijk})\times$$
$$\times\Pi_{ij}^{\alpha}(U_{ni}-U_{nj})
[C^{\mu\nu\sigma\lambda}_{1111}\partial_{R1}^{\sigma}\partial_{R1}^{\lambda}     
+C^{\mu\nu\sigma\lambda}_{1122}\partial_{R2}^{\sigma}\partial_{R2}^{\lambda}$$
$$+C^{\mu\nu\sigma\lambda}_{1133}\partial_{R3}^{\sigma}\partial_{R3}^{\lambda}
+2C^{\mu\nu\sigma\lambda}_{1112}\partial_{R1}^{\sigma}\partial_{R2}^{\lambda}$$
\begin{equation}\label{MFMUEI P time der IV order c}
-2C^{\mu\nu\sigma\lambda}_{1113}\partial_{R1}^{\sigma}\partial_{R3}^{\lambda}
-2C^{\mu\nu\sigma\lambda}_{1123}\partial_{R2}^{\sigma}\partial_{R3}^{\lambda}]
\Psi^{\dagger}\hat{s}_{n}^{\beta}\hat{s}_{j}^{\gamma}\hat{s}_{i}^{\delta}
\Psi dR. \end{equation}
Function $\partial_{t}P^{\alpha}_{(4c)}$ can be considered s two group of terms.
The first three terms give the following structure of the spin densities
$\varepsilon^{\beta\gamma\delta}
[S^{\beta}\cdot S^{\gamma}\cdot \partial^{\sigma}\partial^{\lambda}S^{\delta}]=0$.
Here we used the correlationless limit of the first term,
which is also proportional to $C^{\mu\nu\sigma\lambda}_{1111}$.
Same result appears for other terms in this group.
The second group is constructed of three last terms.
Let us present the tensor structure of the spins in the correlationless limit
$\varepsilon^{\beta\gamma\delta}
[S^{\beta}\cdot \partial^{\sigma}S^{\lambda}\cdot\partial^{\sigma}S^{\delta}]$
(for the term proportional to $C^{\mu\nu\sigma\lambda}_{1112}$).
Tensor structure of the interaction constant is required to make the final conclusion about these terms.
It appears s a superposition of
$\delta^{\mu\nu}\delta^{\sigma\lambda}$,
$\delta^{\mu\sigma}\delta^{\nu\lambda}+\delta^{\mu\lambda}\delta^{\nu\sigma}$,
and
$\delta^{\mu\nu}\delta^{\sigma\lambda}
+\delta^{\mu\sigma}\delta^{\nu\lambda}+\delta^{\mu\lambda}\delta^{\nu\sigma}$.
It leads to the zero value of $\partial_{t}P^{\alpha}_{(4c)}=0$.

The fourth term in $\partial_{t}P^{\alpha}_{(4)}$ appears as  the expansion of the delta-function in the first order
and the expansion of the wave functions up to the third order
$$\partial_{t}P^{\alpha}_{(4d)}=
-\frac{1}{3!}\frac{1}{3^{4}}\varepsilon^{\beta\gamma\delta}
\partial^{\mu}\int \sum_{i,j,i\neq j}\sum_{n\neq i,j}\delta(\textbf{r}-\textbf{R}_{ijk})\times$$

$$\times\Pi_{ij}^{\alpha}(U_{ni}-U_{nj})
[C^{\mu\nu\sigma\lambda}_{1111}\partial_{R1}^{\nu}\partial_{R1}^{\sigma}\partial_{R1}^{\lambda}     
+C^{\mu\nu\sigma\lambda}_{1222}\partial_{R2}^{\nu}\partial_{R2}^{\sigma}\partial_{R2}^{\lambda}$$

$$-C^{\mu\nu\sigma\lambda}_{1333}\partial_{R3}^{\nu}\partial_{R3}^{\sigma}\partial_{R3}^{\lambda}
+3C^{\mu\nu\sigma\lambda}_{1112}\partial_{R1}^{\nu}\partial_{R1}^{\sigma}\partial_{R2}^{\lambda}
-3C^{\mu\nu\sigma\lambda}_{1113}\partial_{R1}^{\nu}\partial_{R1}^{\sigma}\partial_{R3}^{\lambda}$$

$$+3C^{\mu\nu\sigma\lambda}_{1122}\partial_{R1}^{\nu}\partial_{R2}^{\sigma}\partial_{R2}^{\lambda}
+3C^{\mu\nu\sigma\lambda}_{1133}\partial_{R1}^{\nu}\partial_{R3}^{\sigma}\partial_{R3}^{\lambda}
-3C^{\mu\nu\sigma\lambda}_{1223}\partial_{R2}^{\nu}\partial_{R2}^{\sigma}\partial_{R3}^{\lambda}$$

\begin{equation}\label{MFMUEI P time der IV order d}
+3C^{\mu\nu\sigma\lambda}_{1233}\partial_{R2}^{\nu}\partial_{R3}^{\sigma}\partial_{R3}^{\lambda}
-6C^{\mu\nu\sigma\lambda}_{1123}\partial_{R1}^{\nu}\partial_{R2}^{\sigma}\partial_{R3}^{\lambda}]
\Psi^{\dagger}\hat{s}_{n}^{\beta}\hat{s}_{j}^{\gamma}\hat{s}_{i}^{\delta}
\Psi dR. \end{equation}

We can distinguish three types of terms in $\partial_{t}P^{\alpha}_{(4d)}$.
The first group is composed of three first terms in equation (\ref{MFMUEI P time der IV order d}).
So, these are terms proportional to $C^{\mu\nu\sigma\lambda}_{1111}$, $C^{\mu\nu\sigma\lambda}_{1222}$,
and $C^{\mu\nu\sigma\lambda}_{1333}$.
For the term proportional to $C^{\mu\nu\sigma\lambda}_{1333}$
we find the following correlationless limit
$\varepsilon^{\beta\gamma\delta}
[\partial^{\nu}\partial^{\sigma}\partial^{\lambda}S^{\beta}\cdot S^{\gamma}\cdot S^{\delta}]=0$.
Same conclusion we make for two other terms.

Next, we consider the last term in equation (\ref{MFMUEI P time der IV order d}).
Its spin structure has the following form (in the correlationless limit)
$\varepsilon^{\beta\gamma\delta}
\partial^{\lambda}S^{\beta}\cdot \partial^{\sigma}S^{\gamma}\cdot\partial^{\nu} S^{\delta}$.
Tensor structure of the interaction constant appears as the following tensor elements
$\delta^{\mu\lambda}\delta^{\nu\sigma}$,
$\delta^{\mu\nu}\delta^{\sigma\lambda}
+\delta^{\mu\sigma}\delta^{\nu\lambda}$, and
$\delta^{\mu\nu}\delta^{\sigma\lambda}
+\delta^{\mu\sigma}\delta^{\nu\lambda}+\delta^{\mu\lambda}\delta^{\nu\sigma}$.
It leads to the zero value of the considering term.

Finally, we need to consider six terms with numbers four, five, six, seven, eight, nine.
All of them are proportional to number $3$.
The correlationless limit gives the following spin structure
$\varepsilon^{\beta\gamma\delta}
[S^{\beta}\cdot \partial^{\lambda}S^{\gamma}\cdot \partial^{\nu}\partial^{\sigma}S^{\delta}]$.
This tensor structure gives nonzero contribution if coefficient in front of it is nonzero.
Calculation gives coefficient proportional to
$\delta^{\mu\nu}\delta^{\sigma\lambda}
+\delta^{\mu\sigma}\delta^{\nu\lambda}+\delta^{\mu\lambda}\delta^{\nu\sigma}$.

Hence, the last group of terms gives nonzero contribution,
which can be presented in the following form
\begin{equation}\label{MFMUEI P time der IV order d fin simplified}
\partial_{t}P^{\alpha}_{(4d)}=
\frac{1}{3!}\frac{1}{3}\varepsilon^{\beta\gamma\delta}g_{u}g_{\Pi}^{\alpha}
[S^{\beta}\cdot \partial^{\mu}S^{\gamma}\cdot\partial^{\mu}\triangle S^{\delta}]
, \end{equation}
where
$$g_{u}=\int r^{2}U(r)d^{3}r,$$
and
$$g_{\Pi}^{\alpha}=\int r^{2}\Pi^{\alpha}(r)d^{3}r.$$

At the representation of equation (\ref{MFMUEI P time der IV order d fin simplified})
we used
$\varepsilon^{\beta\gamma\delta}[\partial^{\mu}S^{\beta}\cdot \partial^{\mu}S^{\gamma}\cdot\triangle S^{\delta}]=0$
due to the symmetry on $\beta$ and $\gamma$ in the product of the spins,
and
$\varepsilon^{\beta\gamma\delta}[S^{\beta}\cdot \partial^{\mu}\partial^{\mu}S^{\gamma}\cdot\partial^{\mu}\triangle S^{\delta}]=0$
with $\partial^{\mu}\partial^{\mu}=\triangle$,
due to the symmetry on $\gamma$ and $\delta$ in the product of the spins,
$\varepsilon^{\beta\gamma\delta}
[\partial^{\mu}S^{\beta}\cdot \partial^{\nu}S^{\gamma}\cdot\partial^{\mu}\partial^{\nu} S^{\delta}]=0$
due to the following algebra
$\varepsilon^{\beta\gamma\delta}
[\partial^{\mu}S^{\beta}\cdot \partial^{\nu}S^{\gamma}\cdot\partial^{\mu}\partial^{\nu} S^{\delta}]$
$=\varepsilon^{\beta\gamma\delta}
[\partial^{\nu}S^{\beta}\cdot \partial^{\mu}S^{\gamma}\cdot\partial^{\mu}\partial^{\nu} S^{\delta}]$
$=\frac{1}{2}\varepsilon^{\beta\gamma\delta}
[(\partial^{\mu}S^{\beta}\cdot \partial^{\nu}S^{\gamma}$
$+\partial^{\nu}S^{\beta}\cdot \partial^{\mu}S^{\gamma})\cdot\partial^{\mu}\partial^{\nu} S^{\delta}]=0$
due to the symmetry on $\beta$ and $\gamma$ in the product of the spins,
and
$\varepsilon^{\beta\gamma\delta}
[S^{\beta}\cdot \partial^{\mu}\partial^{\nu}S^{\gamma}\cdot\partial^{\mu}\partial^{\nu} S^{\delta}]$
due to the symmetry on $\gamma$ and $\delta$ in the product of the spins.

The fifth and final part of $\partial_{t}P^{\alpha}_{(4)}$ appears at the consideration of the zeroth order expansion of the delta-function
and the expansion of the product of the wave functions up to the fourth order
$$\partial_{t}P^{\alpha}_{(4e)}=
\frac{1}{4!}\frac{1}{3^{4}}\varepsilon^{\beta\gamma\delta}
\int \sum_{i,j,i\neq j}\sum_{n\neq i,j}\delta(\textbf{r}-\textbf{R}_{ijk})\Pi_{ij}^{\alpha}\times$$
$$\times(U_{ni}-U_{nj})
[C^{\mu\nu\sigma\lambda}_{1111}\partial_{R1}^{\mu}\partial_{R1}^{\nu}\partial_{R1}^{\sigma}\partial_{R1}^{\lambda}
+C^{\mu\nu\sigma\lambda}_{2222}\partial_{R2}^{\mu}\partial_{R2}^{\nu}\partial_{R2}^{\sigma}\partial_{R2}^{\lambda}$$
$$+C^{\mu\nu\sigma\lambda}_{3333}\partial_{R3}^{\mu}\partial_{R3}^{\nu}\partial_{R3}^{\sigma}\partial_{R3}^{\lambda}
+4C^{\mu\nu\sigma\lambda}_{1112}\partial_{R1}^{\mu}\partial_{R1}^{\nu}\partial_{R1}^{\sigma}\partial_{R2}^{\lambda}$$

$$
-4C^{\mu\nu\sigma\lambda}_{1113}\partial_{R1}^{\mu}\partial_{R1}^{\nu}\partial_{R1}^{\sigma}\partial_{R3}^{\lambda}
+4C^{\mu\nu\sigma\lambda}_{2221}\partial_{R2}^{\mu}\partial_{R2}^{\nu}\partial_{R2}^{\sigma}\partial_{R1}^{\lambda}$$
$$
-4C^{\mu\nu\sigma\lambda}_{2223}\partial_{R2}^{\mu}\partial_{R2}^{\nu}\partial_{R2}^{\sigma}\partial_{R3}^{\lambda}
-4C^{\mu\nu\sigma\lambda}_{3331}\partial_{R3}^{\mu}\partial_{R3}^{\nu}\partial_{R3}^{\sigma}\partial_{R1}^{\lambda}$$
$$
-4C^{\mu\nu\sigma\lambda}_{3332}\partial_{R3}^{\mu}\partial_{R3}^{\nu}\partial_{R3}^{\sigma}\partial_{R2}^{\lambda}
+6C^{\mu\nu\sigma\lambda}_{1122}\partial_{R1}^{\mu}\partial_{R1}^{\nu}\partial_{R2}^{\sigma}\partial_{R2}^{\lambda}$$

$$
+6C^{\mu\nu\sigma\lambda}_{1133}\partial_{R1}^{\mu}\partial_{R1}^{\nu}\partial_{R3}^{\sigma}\partial_{R3}^{\lambda}
+6C^{\mu\nu\sigma\lambda}_{2233}\partial_{R2}^{\mu}\partial_{R2}^{\nu}\partial_{R3}^{\sigma}\partial_{R3}^{\lambda}$$
$$
-12C^{\mu\nu\sigma\lambda}_{1123}\partial_{R1}^{\mu}\partial_{R1}^{\nu}\partial_{R2}^{\sigma}\partial_{R3}^{\lambda}
-12C^{\mu\nu\sigma\lambda}_{2213}\partial_{R2}^{\mu}\partial_{R2}^{\nu}\partial_{R1}^{\sigma}\partial_{R3}^{\lambda}$$
\begin{equation}\label{MFMUEI P time der IV order e}
+12C^{\mu\nu\sigma\lambda}_{3312}\partial_{R3}^{\mu}\partial_{R3}^{\nu}\partial_{R1}^{\sigma}\partial_{R2}^{\lambda}]
\Psi^{\dagger}\hat{s}_{n}^{\beta}\hat{s}_{j}^{\gamma}\hat{s}_{i}^{\delta}
\Psi dR. \end{equation}
Equation (\ref{MFMUEI P time der IV order e}), in the correlationless limit, gives several group of terms,
which should be considered separately.
Let us consider the first term,
which is proportional to $C^{\mu\nu\sigma\lambda}_{1111}$.
It leads to $\varepsilon^{\beta\gamma\delta}S^{\beta}\cdot S^{\gamma}$$\cdot\partial^{\mu}\partial^{\nu}\partial^{\sigma}\partial^{\lambda} S^{\delta}=0$.
Same result we obtain for the second term $\sim C^{\mu\nu\sigma\lambda}_{2222}$ and the third term $\sim C^{\mu\nu\sigma\lambda}_{3333}$.

Next, let us consider last three terms, each of them is proportional to number $12$.
Considering this group,
we start our analysis with term proportional to $C^{\mu\nu\sigma\lambda}_{1123}$.
In the correlationless limit, it is calculated and represented as follows
$-\frac{12}{4!}\frac{1}{3^{4}}\varepsilon^{\beta\gamma\delta}\hat{g}^{\alpha,\mu\nu\sigma\lambda}_{1123}$
$[\partial^{\lambda}S^{\beta}\cdot \partial^{\sigma}S^{\gamma}\cdot\partial^{\mu}\partial^{\nu} S^{\delta}]$,
where
$\hat{g}^{\alpha,\mu\nu\sigma\lambda}_{1123}$
is a combination of constants
$g_{0u}=\int U(r)d^{3}r$,
$g_{0\Pi}^{\alpha}=\int \Pi^{\alpha}d^{3}r$,
$g_{4u}=\int r^{4}U(r)d^{3}r$,
$g^{\alpha}_{4\Pi}=\int r^{4}\Pi^{\alpha}(r)d^{3}r$,
$g_{u}$, $g_{\Pi}^{\alpha}$
and tensors
$\delta^{\mu\nu}\delta^{\sigma\lambda}$,
$\delta^{\mu\sigma}\delta^{\nu\lambda}+\delta^{\mu\lambda}\delta^{\nu\sigma}$, $I_{0}^{\mu\nu\sigma\lambda}$.
The analysis of the tensor structure of the term proportional to $C^{\mu\nu\sigma\lambda}_{1123}$ shows
that it is equal to zero.
Similar calculations shows that the terms proportional to $C^{\mu\nu\sigma\lambda}_{2213}$ and $C^{\mu\nu\sigma\lambda}_{3312}$ are equal to zero as well.

As the third group of terms we consider terms number ten, eleven, twelve
of equation (\ref{MFMUEI P time der IV order e}).
Each of them is proportional to number $6$.
If we consider term number ten proportional to $C^{\mu\nu\sigma\lambda}_{1122}$
we can show the spin structure appearing in the correlationless limit
$\varepsilon^{\beta\gamma\delta}
[S^{\beta}\cdot \partial^{\sigma}\partial^{\lambda}S^{\gamma}\cdot\partial^{\mu}\partial^{\nu} S^{\delta}]$.
Terms proportional to $C^{\mu\nu\sigma\lambda}_{1133}$ and $C^{\mu\nu\sigma\lambda}_{2233}$ cancel each other.
Calculation of the interaction constant in the term proportional to $C^{\mu\nu\sigma\lambda}_{1122}$ shows
that it is superposition of $I_{0}^{\mu\nu\sigma\lambda}$,
$\delta^{\mu\nu}\delta^{\sigma\lambda}$, and
$\delta^{\mu\sigma}\delta^{\nu\lambda}+\delta^{\mu\lambda}\delta^{\nu\sigma}$.
Together with the spin density structure demonstrated above it gives the zero value.

The final group of terms in equation (\ref{MFMUEI P time der IV order e}) is combined
of six terms number four, five, six, seven, eight, nine.
Each of them is proportional to number $4$.
This group of terms gives nonzero result in contrast to other parts of equation (\ref{MFMUEI P time der IV order e}) described above.
Each term, in the correlationless limit, is proportional to
$\varepsilon^{\beta\gamma\delta}
S^{\beta}\cdot \partial^{\mu}S^{\gamma}\cdot\partial^{\mu}\triangle S^{\delta}$.
To complete calculation we need to find the coefficient in front of this spin structure.
This coefficient appears to be equal to zero.
So, we find $\partial_{t}P^{\alpha}_{(4e)}=0$.

Now, we can combine all partial results of calculation of $\partial_{t}P^{\alpha}_{(4)}$.
It reduces to
$\partial_{t}P^{\alpha}_{(4)}=\partial_{t}P^{\alpha}_{(4d)}$
presented above (\ref{MFMUEI P time der IV order d fin simplified}).


\begin{thebibliography}{17}

\bibitem{Khomskii JETP 21} D. I. Khomskii
"Multiferroics and Beyond: Electric Properties of Different Magnetic Textures",
Journal of Experimental and Theoretical Physics \textbf{132}, 482 (2021).





\bibitem{Katsura PRL 07}
H. Katsura, A. V. Balatsky, N. Nagaosa,
"Dynamical Magnetoelectric Coupling in Helical Magnets",
Phys. Rev. Lett. \textbf{98}, 027203 (2007).


\bibitem{Chen EPJB 13}
H.-B. Chen, Y.-Q. Li,
"Dynamical magnetoelectric effects in the distorted spiral multiferroic magnets",
Eur. Phys. J. B \textbf{86}, 376 (2013).




\bibitem{Fyhn PRB 23}
E. H. Fyhn, A. Brataas, A. Qaiumzadeh, and J. Linder,
"Quasiclassical theory for antiferromagnetic metals",
Phys. Rev. B \textbf{107}, 174503 (2023).

\bibitem{Juraschek PRM 17}
D. M. Juraschek, M. Fechner, A. V. Balatsky, and N. A. Spaldin,
"Dynamical multiferroicity",
Phys. Rev. Materials \textbf{1}, 014401 (2017).



\bibitem{Pimenov PRL 09} A. Pimenov, A. Shuvaev, A. Loidl, F. Schrettle, A. A. Mukhin, V. D. Travkin, V. Yu. Ivanov, and A. M. Balbashov,
"Magnetic and Magnetoelectric Excitations in TbMnO$_{3}$"
Phys. Rev. Lett. \textbf{102}, 107203 (2009).

\bibitem{Chen PRB 16}
Z. Chen, M. Schmidt, Z. Wang, F. Mayr, J. Deisenhofer, A. A. Mukhin, A. M. Balbashov, and A. Loidl,
"Electromagnons, magnons, and phonons in Eu$_{1–x}$Ho$_{x}$MnO$_{3}$",
Phys. Rev. B \textbf{93}, 134406 (2016).


\bibitem{Drofa TMP 96}  M. A. Drofa, L. S. Kuz'menkov,
"Continual approach to multiparticle systems with long-range interaction. Hierarchy of macroscopic fields and physical consequences",
Theoretical and Mathematical Physics \textbf{108}, 849 (1996).



\bibitem{Andreev PRB 11} P. A. Andreev, L. S. Kuzmenkov, M. I. Trukhanova,
"Quantum hydrodynamics approach to the formation of waves in polarized two-dimensional systems
of charged and neutral particles",
Phys. Rev. B \textbf{84}, 245401 (2011).

\bibitem{Andreev EPL 20} P. A. Andreev, M. I. Trukhanova,
"Dipolar drift and collective instabilities of skyrmions
in crossed nonuniform electric and magnetic fields
in a chiral magnetic insulator",
EPL \textbf{132}, 56002 (2020).


\bibitem{Trukhanova Andreev 2305} M. I. Trukhanova, P. Andreev, 
"A New Microscopic Representation of Spin Dynamics in Quantum Systems with Biquadratic Exchange Interactions",
Moscow University Physics Bulletin \textbf{79}, 232 (2024).

\bibitem{Maksimov QHM 99} L. S. Kuz'menkov, S. G. Maksimov, "Quantum hydrodynamics of particle systems with
coulomb interaction and quantum bohm potential," Theor. Math. Phys. \textbf{118}, 227 (1999).


\bibitem{MaksimovTMP 2001} L. S. Kuz'menkov, S. G. Maksimov, and V. V. Fedoseev, "Microscopic quantum hydrodynamics of systems of
fermions: Part I," Theoretical and Mathematical
Physics \textbf{126}, 110 (2001).



\bibitem{Andreev Ch 21} P. A. Andreev, "Quantum hydrodynamic theory of quantum fluctuations in dipolar Bose–-Einstein condensate"
Chaos \textbf{31}, 023120 (2021).


\bibitem{Andreev PoF 21} P. A. Andreev, I. N. Mosaki, and M. I. Trukhanova,
"Quantum hydrodynamics of the spinor Bose–Einstein condensate at non-zero temperatures",
Phys. Fluids \textbf{33}, 067108 (2021).


\bibitem{Andreev PTEP 19 spin current}
P. A. Andreev, L. S. Kuz'menkov,
"On the equation of state for the "thermal" part of the spin current:
The Pauli principle contribution in the spin wave spectrum in a cold fermion system"
Prog. Theor. Exp. Phys. \textbf{2019}, 053J01 (2019).



\bibitem{Shukla RMP 11} P. K. Shukla, B. Eliasson,
"Nonlinear collective interactions in quantum plasmas with degenerate electron fluids",
Rev. Mod. Phys. \textbf{83}, 885 (2011).


\bibitem{Wigner PR 84} M. Hillery, R. F. O'Connell, M. O. Scully, E. P. Wigner,
"Distribution functions in physics: Fundamentals",
Physics Reports \textbf{106}, 121 (1984).

\bibitem{Silin ZETF 85}
V. P. Silin and A. Z. Solontsov,
"Theory of the temperature dependence of the magnon spectrum in ferromagnetic metals",
Zh. Eksp. Teor. Fiz. \textbf{89}, 1432 (1985).


\bibitem{Kondratyev LTP 08}
A. S. Kondratyev, I. Siddique,
"Kinetic equations in the theory of normal Fermi liquid",
Low Temp. Phys. \textbf{34}, 137 (2008).




\bibitem{Andreev LP 19} P. A. Andreev,
"Hydrodynamic model of a Bose–-Einstein condensate with anisotropic
short-range interaction and bright solitons in a repulsive Bose–Einstein condensate",
Laser Phys. \textbf{29}, 035502 (2019).


\bibitem{Andreev LP 21} P. A. Andreev,
"Extended hydrodynamics of the degenerate partially spin polarized fermions
with the short-range interaction up to the third order by the interaction radius approximation",
Laser Phys. \textbf{31}, 045501 (2021).
%








\bibitem{Tokura RPP 14} Y. Tokura, S. Seki, and N. Nagaosa,
"Multiferroics of spin origin",
Rep. Prog. Phys. \textbf{77}, 076501 (2014).





%
\bibitem{Kawaguchi Ph Rep 12} Y. Kawaguchi, M. Ueda,
"Theory of spin-2 Bose-Einstein condensates: Spin correlations, magnetic response, and excitation spectra",
Phys. Rep. \textbf{520},  253 (2012).

\bibitem{Stamper-Kurn RMP 13} D. M. Stamper-Kurn, M. Ueda,
"Spinor Bose-Einstein condensates",
Rev. Mod. Phys. \textbf{85}, 1191 (2013).
%





%



\bibitem{Barman JAP 20} A. Barman, S. Mondal, S. Sahoo, and A. De
"Magnetization dynamics of nanoscale magnetic materials: A perspective",
J. Appl. Phys. \textbf{128}, 170901 (2020).


\end{thebibliography}
\end{document}